\documentstyle[aps,version2,preprint]{revtex}    
\begin{document}

\draft

\begin{title}
New perturbation theory of low-dimensional quantum liquids I:\\
the pseudoparticle operator basis
\end{title}

\author{J.M.P. Carmelo $^{1,2,*}$, A.H. Castro Neto
$^{2}$, and D.K. Campbell $^{2}$}
\begin{instit}
$^{1}$ Instituto de Ciencia de Materiales, C.S.I.C.,
Cantoblanco, SP - 28949 Madrid, Spain
\end{instit}
\begin{instit}
$^{2}$ Department of Physics, University of Illinois at
Urbana -- Champaign, \\
1110 West Green Street, Urbana, Illinois 61801-3080
\end{instit}
\receipt{February 1994}

\begin{abstract}
We introduce a new operator algebra for the description of
the low-energy physics of one-dimensional,
integrable, multicomponent quantum liquids. Considering
the particular case of the Hubbard chain in a constant
external magnetic field and with varying
chemical potential, we
show that at low energy its Bethe-ansatz
solution can be interpreted in terms of
the new {\it pseudoparticle operator algebra}.
Our algebraic approach provides a concise interpretation
of and justification for several recent studies of
low-energy excitations and transport which have been
based on detailed analyses of specific Bethe-ansatz
eigenfunctions and eigenenergies.
A central point is that the {\it exact ground state} of the
interacting many-electron problem is the non-interacting pseudoparticle ground
state. Furthermore, in the pseudoparticle basis, the
quantum problem becomes perturbative, {\it i.e.}, the
two-pseudoparticle forward-scattering vertices and amplitudes
do not diverge, and one can define a many-pseudoparticle
perturbation theory. We write the general quantum-liquid
Hamiltonian in the new basis and show
that the pseudoparticle-perturbation theory leads, in a
natural way, to the generalized Landau-liquid approach.
\end{abstract}
\renewcommand{\baselinestretch}{1.656}   


\narrowtext

\section{INTRODUCTION}

For more than sixty years the ``Bethe ansatz'' (BA) \cite{Bethe,Yang}
has played a central role in the analytic solution of a variety of
``integrable''
\cite{Thacker} many-body problems in condensed matter physics and
quantum field theory \cite{Bethe,Yang,Thacker,Lieb,Korepinrev}. Initially
applied to determine the ground state energies and spatial eigenfunctions,
BA techniques were generalized to describe excited states, thermodynamics,
and correlation functions \cite{Thacker,Korepinrev} and remain
an active subject of study today. Spurred by a conjectured
relationship to the microscopic mechanism for high-temperature
superconductivity \cite{Ander}, those particular integrable interacting
electron models -- such as the Hubbard Hamiltonian \cite{Lieb} --
which exhibit ``Luttinger-liquid'' behavior \cite{Haldane} have
been extensively investigated
\cite{Izergin,Woy,Parola,Schulz,Ren,Frahm,Carmelo92,Carmelo92c}.

In particular, recent investigations have established the existence
of Hubbard-model eigenstates possessing off-diagonal order
(a feature essential for superconductivity) \cite{Yang89}
and have shown that, although not themselves complete, the BA
eigenstates can be extended to form a complete set of states for
the one-dimensional Hubbard model \cite{Korepin}. Importantly, both
these new results have been derived using powerful algebraic techniques,
which rely on the symmetries of the Hubbard Hamiltonian. A vital
and to date open issue is the extent to which these
or similar algebraic
techniques and symmetries can be used to describe specifically those
excitations that dominate response and transport at low energies in
integrable quantum liquids and related models.

In the present article, which extends our earlier results \cite{Neto},
we show that there is a new operator
algebra associated with the low-energy Hamiltonian eigenstates
of integrable quantum liquids solvable by BA
\cite{Bethe,Yang,Lieb,Izergin,Frahm}.
We show that this algebra is expressed most naturally in terms
of operators describing the ``pseudoparticle'' excitations
introduced in several previous papers
\cite{Carmelo92,Carmelo92c,Neto,Carmelo91,Carmelo91b,Carmelo92b,Campbell}
and proven to dominate low-energy transport and response functions.

Whereas this previous work focused on the eigenstates
and eigenfunctions of the pseudoparticles, here we are able to
work in terms of algebraic operators alone and hence to obtain
a much more compact and general representation.
Considering the particular case of the Hubbard chain in a magnetic
field and chemical potential, we demonstrate that the
familiar BA solution can be interpreted naturally in terms of the
pseudoparticle basis and that the Hamiltonian can be expressed simply
in terms of operators in that basis. We then establish that a pseudoparticle
perturbation theory, which is also naturally described in
terms of the new operator algebra, can be used to study the low-energy
excitations and the response and correlation functions.

Our primary goal in the present article is to establish the utility
of the pseudoparticle operator algebra for expressing the
Hamiltonian and for calculating in perturbation
theory the low-energy excitations of the one-dimensional Hubbard and
related models. In a companion paper,
which we shall henceforth call II (\cite{Neto93}), we
will apply the perturbation theory introduced here
to the study of the Virasoro
operator algebras of multicomponent integrable quantum
liquids. Papers I and II contain a detailed exposition and
extension of our earlier results \cite{Neto}.

In Section II we introduce the Hubbard Hamiltonian in the
presence of non-zero external magnetic field and chemical
potential. We discuss prior results establishing that in this
case the low-energy physics is dominated by
a particular class of eigenstates:
these lowest-weight states (LWS) of
both the $\eta$ spin and spin algebras
\cite{Yang89,Korepin,Heilmann,Lieb89,Zhang,Ostlund,Essler,Two,Nuno}
which refer to real rapidities \cite{Carmelo92c}. We call these
``LWS I'' to distinguish them from the LWS associated with
complex, non-real, rapidities, which we call ``LWS II''. In
the sectors of parameter space with $U(1)\times U(1)$
symmetry the LWS II excitations have an energy gap
and do not contribute to the low-energy physics.
The LWS I can be described simply in terms of the pseudoparticles and
the corresponding algebra. We demonstrate that in the
pseudoparticle basis the ground state of the many-body problem is a
``non-interacting'' state, {\it i.e.}, a simple Slater
determinant of filled pseudoparticle levels. A
central point is that for all canonical ensembles
and symmetries with densities $0\leq n\leq 1$ and spin
densities $0\leq m\leq n$ the ground state is {\it always}
an LWS I. We discuss the
nature of the pseudoparticle Fermi sea and the
low-energy excitations, showing that these
excitations can be simply understood as pseudoparticle-pseudohole
pairs and that they span the full low-energy Hilbert space.
We introduce pseudomomentum-density operators and show that the
operator expressions for the momentum, pseudoparticle number, and
pseudoparticle-density fluctuations all have the expected forms.

In Section III we relate the important rapidity numbers used
in the BA to the pseudomomentum-distribution operators and use this
relation to write the Hamiltonian (formally)
in terms of the pseudoparticle operators.

In Section IV we discuss the (potentially important)
effects of normal-ordering and show that, in terms of the
pseudoparticle operators, the normal-ordered Hamiltonian
in the low-energy subspace has a simple form. This permits
the introduction of a systematic, non-singular pseudoparticle
perturbation theory. Further, it establishes directly the {\it universal}
form of the Hamiltonian for integrable multicomponent
quantum liquids. These two results lead, in a natural way, to the
one-dimensional Landau-liquid theory studied in Refs.
\cite{Carmelo92,Carmelo92c,Neto,Carmelo91,Carmelo91b,Carmelo92b,Campbell}.
Our operator representation  normal-ordered with respect to
the ground state leads also to the results obtained
from conformal-field theory  \cite{Frahm,Neto}, as we show
in detail in paper II: the energy-momentum
stress tensor and the generators of the Virasoro algebras
can also be written simply in the pseudoparticle operator
representation.

In Section V we present a discussion and concluding remarks.
We indicate how results derived for the Hubbard
model can be extended to
multicomponent BA solvable models \cite{Izergin,Frahm,Neto} in which the
theory has $\nu\geq 2$ independent branches of
gapless elementary excitations.
We contrast the perturbative character of the
interaction effects in the pseudoparticle approach
with the non-perturbative character of the usual
electronic basis. We explain that the universal character
of the present class of one-dimensional integrable quantum
liquids can be understood in terms of a straightforward
generalization to pseudoparticles of Wilson's
renormalization group arguments \cite{Castro}: the pseudo-Fermi points
of the pseudoparticles replace the particle Fermi surface and
close to the pseudo-Fermi points only a few
two-pseudoparticle scattering processes are relevant
for the low-energy physics. Finally, we indicate briefly possible
implications of our results for real quasi-one dimensional
materials and for two-dimensional quantum systems.

\section{THE PSEUDOPARTICLE OPERATOR BASIS}

Consider the Hamiltonian describing the Hubbard chain
in an external magnetic field $H$ and with chemical potential $\mu$
\cite{Frahm,Carmelo92c,Carmelo91b,Carmelo92b}:

\begin{equation}
\hat{H} = -t\sum_{j,\sigma}\left[c_{j\sigma}^{\dag }c_{j+1\sigma}+c_
{j+1\sigma}^{\dag }c_{j\sigma}\right] +
U\sum_{j} [c_{j\uparrow}^{\dag }c_{j\uparrow} - 1/2]
[c_{j\downarrow}^{\dag }c_{j\downarrow} - 1/2]
- 2\mu\hat{\eta }_z - 2\mu_0 H\hat{S}_z \, ,
\label{Hamilt}
\end{equation}
where $t$, $U$, $\mu$, $H$, and $\mu _0$ are the transfer
integral, the onsite ``Coulomb'' interaction, the chemical potential,
the magnetic field, and the Bohr magneton, respectively. The
operator $c_{j\sigma}^{\dagger}$ ($c_{j\sigma}$) creates (annihilates)
a spin $\sigma$ electron at the site $j$, and

\begin{equation}
\hat{\eta}_z = -{1\over 2}[N_a - \sum_{\sigma}\hat{N}_{\sigma }] \, ,
\hspace{1cm}
\hat{S}_z = -{1\over 2}\sum_{\sigma}\sigma\hat{N}_{\sigma } \, ,
\end{equation}
are the diagonal generators of the $SU(2)$ $\eta$ spin
and spin algebras, respectively \cite{Yang89,Korepin,Two,Nuno}.
As usual, $\sigma$ refers to the spin projections $\sigma =\uparrow\,
,\downarrow$ when used as an operator or function index
and is given by $\sigma =\pm 1$ otherwise. In Eqs.
$(2)$

\begin{equation}
\hat{N}_{\sigma } = \sum_{j}
c_{j\sigma}^{\dagger }c_{j\sigma}
= \sum_{k'}
c^{\dagger }_{k'\sigma} c_{k'\sigma}
= \hat{\rho}^{\sigma }(0) \, , \hspace{1cm}
\hat{\rho}^{\sigma}(k) \equiv \sum_{k'}
c^{\dagger }_{k'+k\sigma} c_{k'\sigma} \, ,
\end{equation}
is the number operator for spin $\sigma$ electrons. The operator
$\hat{\rho}^{\sigma}(k)$ in Eq. $(3)$
is the one-pair electron operator associated with
the spin $\sigma$ fluctuations. Here $c^{\dagger}_{k\sigma}$
($c_{k\sigma}$) is the momentum Fourier representation of the
operator $c_{j\sigma}^{\dagger}$ ($c_{j\sigma}$).

The one-dimensional Hubbard Hamiltonian $(1)$ describes an
interacting quantum system of $N_{\downarrow}$ down-spin
electrons and $N_{\uparrow}$ up-spin electrons
on a chain of $N_a$ sites with lattice
constant $a$. Henceforth we employ units such that
$a=t=\mu_0=\hbar =1$. Introducing $n_{\sigma} = N_{\sigma}/N_a$
and  $n = n_{\downarrow} + n_{\uparrow}$,
we define ($k_{F\sigma}
=\pi n_{\sigma}$) and the total Fermi momentum by
$k_F=\pi n/2$.

In the absence of the chemical-potential and magnetic-field terms
the Hamiltonian $(1)$ has $SO(4) = SU(2) \otimes SU(2)/Z_2$ symmetry
\cite{Yang89,Korepin,Heilmann,Lieb89,Zhang,Essler,Two,Nuno}.
When $N_a$ is even, the operator ${\hat{\eta}}_z
+ {\hat{S}}_z$ (see Eq. $(2)$) has only integer eigenvalues
and all half-odd integer representations of
$SU(2) \otimes SU(2)$ are projected out
\cite{Korepin,Essler}.The two $SU(2)$ algebras -- $\eta$ spin
and spin  -- have diagonal generators given by Eq. (2) and
off-diagonal generators \cite{Korepin,Essler}

\begin{equation}
\hat{\eta} = \sum_{j} (-1)^j c_{j\uparrow}c_{j\downarrow}
\, , \hspace{1cm}
\hat{\eta}^{\dagger} = \sum_{j} (-1)^j c^{\dagger }_{j\downarrow}
c^{\dagger }_{j\uparrow} \, ,
\end{equation}
and

\begin{equation}
\hat{S} = \sum_{j} c^{\dagger }_{j\uparrow}c_{j\downarrow}
\, , \hspace{1cm}
\hat{S}^{\dagger} = \sum_{j} c^{\dagger }_{j\downarrow}
c_{j\uparrow} \, ,
\end{equation}
respectively.

In the presence of both the magnetic field and chemical
potential terms, the symmetry is reduced to $U(1)\otimes U(1)$,
with $\hat{\eta}_z$ and $\hat{S}_z$ commuting with $\hat{H}$.
Hence at any fixed value of $U$, the parameter space
of the Hamiltonian $(1)$ is two dimensional
($\nu$ dimensional for the multicomponent case
when the number of subalgebras
is $\nu$ and the symmetry is $[U(1)]^{\nu}$).
According to these values, the system has different
symmetries as follows \cite{Two}: when $\eta_z\neq 0$ and
$S_z\neq 0$ the symmetry is $U(1)\otimes U(1)$, for $\eta_z= 0$
and $S_z\neq 0$ it is $SU(2)\otimes U(1)$, when $\eta_z\neq 0$ and
$S_z= 0$ it is $U(1)\otimes SU(2)$, and at $\eta_z= 0$ and $S_z= 0$
the symmetry is $SO(4)$. Note
that the $U(1)\otimes U(1)$ symmetry sector {\it always}
corresponds to two non-zero eigenvalues of
the diagonal generators, whereas in sectors
of higher symmetry,  one or both of these
eigenvalues vanish.

Although the two conserved quantum numbers are usual
taken (as in Eq.$(3)$) to be $\hat{N}_{\uparrow }$ and
$\hat{N}_{\downarrow }$, an alternative choice
is $\hat{N}_{\rho }$ and $\hat{N}_{\sigma_z}$, {\it i.e.}
$\hat{N} = \hat{N}_{\uparrow}+
\hat{N}_{\downarrow}=\hat{\rho}^{\rho}(0)$ and
$\hat{N}_{\sigma_z}=\hat{N}_{\uparrow} -
\hat{N}_{\downarrow}=\hat{\rho}^{\sigma_z }(0)$,
where the charge and spin one-pair electron
operators read

\begin{equation}
\hat{\rho}^{\rho}(k) =
\hat{\rho}^{\uparrow}(k) + \hat{\rho}^{\downarrow}(k)  \, ,
\hspace{2cm}
\hat{\rho}^{\sigma_z }(k) =
\hat{\rho}^{\uparrow}(k) - \hat{\rho}^{\downarrow}(k)  \, .
\end{equation}

The eigenvalues ${\eta}_z$ and $S_z$ can be expressed in terms
of the conserved numbers, as shown by
Eq.$(2)$. This holds true also for
the $\nu$ eigenvalues of the diagonal generators
of $\nu$ multicomponent integrable systems.
In these systems, in addition to the chemical potential and magnetic field,
the system can include other external fields associated
with the additional conserved quantum numbers. Note, however,
that the number $\nu$ of fields always equals the number
of types of particles of the quantum liquid.

In the above discussion we have worked in the parameter
space specified by eigenvalues $({\eta}_z,S_z)$, corresponding
to the canonical ensemble. Equivalently,
we could choose the two external ``fields'' -- $H$ and $\mu$ --
as parameters, or in fact use a ``mixed'' representation.
We remark that in the
papers \cite{Carmelo92,Carmelo92c,Carmelo91b,Carmelo92b}
a mixed representation involving the magnetic field $H$ and
the density $n$ (i.e. number of particles $N$) was often used.

Historically, most investigations of the Hubbard chain and
related models have considered the {\it higher-symmetry sectors}. For instance,
in the case of the Hubbard model studies, have typically
focused on the zero magnetic field ($H=0$) case of symmetry $U(1)\otimes SU(2)$
(or $SO(4)$). By considering the critical
point of the one-dimensional Hubbard model in a magnetic
field, Frahm and Korepin \cite{Frahm} were able to discover
interesting new features that remained hidden in previous
studies of the model at zero magnetic field.
For instance, in contrast to the $H=0$ case, where
the spinon gapless excitations are LWS II, at $H>0$
(and $n\neq 1$) {\it all} gapless Hamiltonian eigenstates
are LWS I and the form of the corresponding BA equations is
much simpler. (The BA equations associated with the
LWS II are in large number and very complicated -- see, for
example, Refs. \cite{Korepin,Essler} -- whereas the
BA equations which describe the LWS I are only two
and have the same structure as the ground-state
equations of Lieb and Wu \cite{Lieb,Frahm,Carmelo92c}.)
However, as discussed in detail in Ref. \cite{Carmelo92c},
the zero magnetic-field limit of the $H>0$ low-energy
expressions provides the correct $H=0$ exponents \cite{Frahm}
and low-energy quantities \cite{Carmelo92c}.
Similarly, several papers focusing on the
lowest symmetry region of parameter space have developed
a pseudoparticle description and studied its relation to
the Landau-liquid properties of the low-energy physics
\cite{Carmelo92,Carmelo92c,Neto,Carmelo91,Carmelo91b,Carmelo92b,Campbell,Neto93}.

We consider here the $U(1)\otimes U(1)$ sector which refers
to electronic densities $0<n<1$ and spin densities
$0<m<n$. This sector contains the ``regular'' BA
states which are LWS of both algebras \cite{Korepin}.
We concentrate our attention on the LWS I.
In this $U(1)\otimes U(1)$ sector
$SO(4)$ multiplets are generated by acting with the raising
operators $(4)$ and $(5)$ on the corresponding LWS
(for both LWS I and II). The total number of states
is $4^{N_a}$ \cite{Korepin}.

It is important to reiterate that our focus on the
$U(1)\otimes U(1)$ symmetry sector, in which
the gapless BA Hamiltonian eigenstates are easier
to describe, does not prevent us from obtaining
relevant information about the sectors of higher symmetry.
This is because, as mentioned above,
the limits of the expressions derived for the
physical quantities in the lowest-symmetry parameter space generally
approach the correct values for the same quantities in the
sectors of higher symmetry \cite{Carmelo92c}.
When the transition from the sector of
parameter space of lowest symmetry to the higher-symmetry
sector does not involve the opening of a gap in one of the
two elementary excitation branches, the values in the higher
symmetry sector are obtained
by taking in the expressions obtained in the $U(1)\otimes U(1)$
sector the limiting values for the eigenvalues of the
diagonal generators which characterize the higher-symmetry
sectors. For example, taking the limit $S_z\rightarrow 0$ in
the expressions for the physical quantities derived in the
present $U(1)\otimes U(1)$ sector leads to the
correct $S_z=0$ expressions of the $U(1)\otimes SU(2)$
(or $SO(4)$) sector. Although the metal-insulator transition
can cause some subtleties  \cite{Carmelo91b,Carmelo92b,Two} in
determining the $n=1$ expressions -- corresponding
to the symmetry $SU(2)\otimes U(1)$ (or $SO(4)$) --  these can also be
obtained from the $\eta_z\rightarrow 0$ values.

As noted above, in the general $\nu$ component case, in the
sectors of lowest symmetry
$[U(1)]^{\nu}$ the number $\nu$ of $U(1)$ subalgebras
also provides the number of independent gapless branches
of elementary excitations \cite{Neto,Two}. As we discuss below,
each of these $\nu$ branches is characterized by a different
``color'' quantum number, which we shall call $\alpha$.
These colors label the pseudoparticle
operators and define ``new'' orthogonal {\it directions} in the
low-energy Hilbert space. Furthermore, the numbers of
pseudoparticles and pseudoholes {\it are} good
quantum numbers. Therefore, at constant values of the
particle-particle interaction $U$, we may use the $\nu$ numbers of $\alpha$
pseudoparticles or the $\nu$ numbers of $\alpha$
pseudoholes as alternatives to the
$\nu$ particle numbers $N_{\sigma}$ associated with
the operators $(3)$, to define the parameter space.

We return now to the specific case of the Hubbard model
and study in detail the operator algebra introduced in
Ref. \cite{Neto}, which generates the Hilbert sub-space spanned
by the regular BA eigenstates described by {\it real}
rapidities (LWS I).

A central point is that in canonical ensembles of
the $U(1)\otimes U(1)$ sector both the non-LWS multiplets
and the LWS II have energy gaps relative
to the ground state \cite{Carmelo92c,Neto,Two}.
Therefore, for energies smaller
than those gaps, the  Hilbert subspace
spanned by the regular BA LWS I {\it coincides}
with the full Hilbert space of the quantum problem.
The perturbation theory which we introduce in Sec. IV refers
to that Hilbert space corresponding to energy scales
{\it smaller} than the above gaps. We note, {\it however},
that LWS I contain {\it both} high- and low-energy
states. In this section we study the entire
Hilbert subspace spanned by the LWS I, but stress
that it coincides with relevant, low-energy
Hilbert space of the quantum problem {\it only} at
energy scales smaller than the above gaps.

The Hilbert subspace spanned by the LWS I can be generated by
acting on the vacuum state with the pseudoparticle algebra we introduce
below. The LWS I pseudoparticles are massless in the
sense of field theory (describe gapless modes); of course, they
have finite static and transport
``condensed-matter'' masses \cite{Carmelo92c,Neto},
which are given in II (for the transport masses see Eq. $(71)$
of II.) For completeness we note that a
more general Landau-liquid theory including the LWS II
states can be constructed \cite{Nuno94}. These LWS can also be
described by pseudoparticles, but these new
pseudoparticles have a {\it mass} in the field-theory
sense (gap, in the present condensed-matter language).

Although the pseudoparticles associated with
the LWS I are the transport carriers at low
energy \cite{Carmelo92c}, they refer to purely
non-dissipative excitations, {\it i.e.} the Hamiltonian
{\it commutes} with the currents in the subspace
spanned by the LWS I \cite{Carmelo92c,Horsch94}. Therefore,
these pseudoparticle currents give rise {\it only}
to the coherent part of the conductivity
spectrum, i.e. to the Drude peak \cite{Carmelo92c,Neto,Neto93}.
The non-coherent part is associated with the
massive pseudoparticles which describe
the LWS II and some of the non-LWS multiplets \cite{Nuno94}.
In the Hilbert subspace spanned by those excitations,
the Hamiltonian {\it does not} commute with the
current operators. In addition, and
to make life more difficult (and interesting),
the $c$ ($s$) excited LWS I states ( $(18)$ below) disappear at $n=1$
(or $H=0$) and the corresponding massive pseudoparticle
becomes massless ($U(1)$ to $SU(2)$ transition)
\cite{Carmelo92c,Nuno94,Horsch94}. Fortunately, for most
of our discussion, this subtlety will be irrelevant.

The new operator algebra studied in this paper
generates {\it all} the LWS I from the electronic vacuum $|V\rangle$.
At constant values of $U$ and $S_z<0$, the vacuum $|V\rangle$ corresponds to
the limit of vanishing electronic density $n\rightarrow 0$.
This is the $S_z < 0$, $U(1)\otimes U(1)$ vacuum which refers
to the limits $n_{\uparrow}\rightarrow 0$ and
$n_{\downarrow}\rightarrow 0$ with $n_{\downarrow}/n_{\uparrow}>1$.
We note that when this vacuum is obtained from the ground
state (given in Eq. $(11)$ below) by taking that limit, in the
case of some matrix elements this does not commute with the limit
$n_{\uparrow}\rightarrow 0$ and $n_{\downarrow}\rightarrow 0$
with $n_{\downarrow}/n_{\uparrow}=1$, which defines
the $S_z = 0$, $U(1)\otimes SU(2)$, vacuum. In turn,
these two limits lead to the same $S_z = 0$ vacuum
when they refer to the ground states alone (see
Eq. $(11)$ below), {\it i.e.} not
to the limiting value of matrix elements involving these states.
This is revealed by the study of the electron - pseudoparticle
canonical transformation which will be presented elsewhere
\cite{Carmelo94}.

The new operator algebra involves two types of {\it pseudoparticle}
creation (annihilation) operators
$b^{\dag }_{q\alpha }$ ($b_{q\alpha }$) which obey the usual
fermionic algebra \cite{Neto}

\begin{equation}
\{b^{\dag }_{q\alpha},b_{q'\alpha'}\}
=\delta_{q,q'}\delta_{\alpha ,\alpha'}, \hspace{0.5cm}
\{b^{\dag }_{q\alpha},b^{\dag }_{q'\alpha'}\}=0, \hspace{0.5cm}
\{b_{q\alpha},b_{q'\alpha'}\}=0 \, .
\end{equation}
Here $\alpha$ refers to the two (or $\nu$, in the general case) pseudoparticle
colors $c$ and $s$ \cite{Carmelo92c,Neto,Two}. The discrete
pseudomomentum values are

\begin{equation}
q_j = {2\pi\over {N_a}}I_j^{\alpha } \, ,
\end{equation}
where $I_j^{\alpha }$ are {\it consecutive} integers or half
integers. (In the case of the Hubbard chain we have that $I_j^c=I_j$
and $I_j^s=J_{\beta}$ [with $\beta=j$], where $I_j$ and
$J_{\beta}$ refers to the notation of Lieb and Wu \cite{Lieb}.)
There are $N_{\alpha }^*$ values of $I_j^{\alpha }$, {\it i.e.}
$j=1,...,N_{\alpha }^*$. An LWS I is specified by
the distribution of $N_{\alpha }$
occupied values, which we call $\alpha $ pseudoparticles, over
the $N_{\alpha }^*$ available values. There are $N_{\alpha }^*-
N_{\alpha }$ corresponding empty values, which we call $\alpha $
pseudoholes. We emphasize that the BA wave functions
vanish for double-occupied configurations of the $I_j^{\alpha }$
quantum numbers \cite{Korepinrev}.

In the present $U(1)\otimes U(1)$ sector of parameter space,
the numbers of $\alpha$ pseudoholes
$N_{\alpha }^*-N_{\alpha }$, such that $N_{\alpha }^*>N_{\alpha }$,
are determined by the corresponding eigenvalues
of the two (or $\nu$) diagonal generators. In the case of the
Hubbard chain we have that

\begin{equation}
-\eta_z = {1\over 2}[N_c^*-N_c] \, ,
\hspace{1cm}
-S_z = {1\over 2}[N_s^*-N_s] \, ,
\end{equation}
where

\begin{equation}
N_c^* = N_a \, ; \hspace{0.5cm}
N_c = N \, ; \hspace{0.5cm}
N_s^* = N_{\uparrow} \, ; \hspace{0.5cm}
N_s = N_{\downarrow} \, .
\end{equation}

The numbers $I_j^c$ are integers (or half integers) for
$N_s$ even (or odd), and $I_j^s$ are integers (or
half integers) for $N_s^*$ odd (or even) \cite{Lieb}. Since only
single and zero occupancy of the values $I_j^{\alpha}$ are allowed,
only pseudoparticles of the color $\alpha$ can occupy
the states labeled by the numbers $I_j^{\alpha}$. Therefore,
the pseudoparticles have a fermionic character, as assured by
the anticommuting algebra $(7)$.
For the Hubbard model the BA spatial wave function
for the LWS I depends on the quantum numbers $I_j^{\alpha}$
through two sets of real numbers, which many authors call rapidities.
Note that for the LWS II these rapidities are complex,
non-real, numbers. The expression of the spatial wave functions
in terms of the quantum numbers $I_j^{\alpha}$ requires the
solution of two systems of algebraic equations which define
the two rapidities as functions of the
quantum numbers $I_j^{\alpha}$
\cite{Yang,Lieb,Frahm,Carmelo92b}.
Although the expression of the spatial wave function for the
LWS I in terms of the quantum numbers
$I_j^{\alpha}$ requires the solution of the above
systems of equations, which constitutes a problem of
considerable complexity, the description of these eigenstates
in the basis of the above BA quantum numbers $I_j^{\alpha}$
that diagonalize the quantum liquid is much simpler.
Historically, the Hamiltonian eigenstates were introduced
in terms of the spatial BA wave functions
\cite{Bethe,Yang}. The diagonalization of the problem leads
then to the two systems of algebraic equations. In the
case of LWS I these equations introduce the integer or
half-integer quantum numbers $I_j^{\alpha}$ which describe
these eigenstates.

One of the principal advantages of the algebraic approach
we use in this paper is that it permits the
description of the LWS I in terms of the quantum numbers
$I_j^{\alpha}$ and does not require the
spatial wave-function representation. In the basis associated
with these quantum numbers the description of the LWS
I does not involve the rapidity numbers.
As in the case of the spatial wave functions, the expression of
the energy in terms of the quantum numbers $I_j^{\alpha}$
involves the rapidities. In Sec. III and Appendix A we will
consider the above algebraic equations for the case of the LWS I of
the Hubbard chain. The introduction of a suitable operator
representation reveals that the rapidities are
the eigenvalues of rapidity operators which in the
pseudoparticle basis determine the Hamiltonian
interaction many-pseudoparticle terms.

The above electronic vacuum state $|V\rangle$ has no electrons.
At fixed values of the onsite interaction $U$, the ground
state of a canonical ensemble
characterized by eigenvalues $\eta_z<0$ and $S_z<0$ can
be constructed from that vacuum. Following Refs.
\cite{Carmelo92c,Neto,Carmelo91b,Carmelo92b,Two} and as shown
elsewhere \cite{Nuno,Nuno94}, it corresponds to filling
symmetrically around the origin $N_{\alpha}$
consecutive $I_j^{\alpha}$ values of all colors $\alpha$.
In addition, the ground state of canonical ensembles
of all symmetries and such that $0\leq n\leq 1$ and $0\leq m\leq n$
are {\it always} LWS I. This holds even at the $SO(4)$ ``point'',
where $n=1$ and $m=0$, and the available Hilbert space
contains {\it only} one LWS I, which is the corresponding
ground state; in this case, all LWS excited states
are LWS II, as we discuss below.
The ground state associated with a canonical
ensemble of $(\eta_z,S_z)$ values has the form

\begin{equation}
|0;\eta_z,S_z\rangle = \prod_{\alpha=c,s}
[\prod_{q=q_{F\alpha }^{(-)}}^{q_{F\alpha }^{(+)}}
b^{\dag }_{q\alpha }]
|V\rangle \, ,
\end{equation}
where when $N_{\alpha }$ is odd (even) and
$I_j^{\alpha }$ are integers (half
integers) the pseudo-Fermi points are symmetric and
given by

\begin{equation}
q_{F\alpha }^{(+)}=-q_{F\alpha }^{(-)} =
{\pi\over {N_a}}[N_{\alpha}-1] \, .
\end{equation}
On the other hand, when $N_{\alpha }$ is odd (even) and
$I_j^{\alpha }$ are half integers (integers)
we have that

\begin{equation}
q_{F\alpha }^{(+)} = {\pi\over {N_a}}N_{\alpha }
\, , \hspace{1cm}
-q_{F\alpha }^{(-)} ={\pi\over {N_a}}[N_{\alpha }-2] \, ,
\end{equation}
or

\begin{equation}
q_{F\alpha }^{(+)} = {\pi\over {N_a}}[N_{\alpha }-2]
\, , \hspace{1cm}
-q_{F\alpha }^{(-)} = {\pi\over {N_a}}N_{\alpha } \, .
\end{equation}
Similar expressions are obtained for the pseudo-Brioullin
zones limits $q_{\alpha }^{(\pm)}$ if we replace
in Eqs. $(12)-(14)$ $N_{\alpha }$ by $N_{\alpha }^*$.

The simple form of the ground-state expression $(11)$ has a
deep physical meaning. It reveals that in the pseudoparticle
basis the ground state of the many-electron quantum problem
is a ``non-interacting'' pseudoparticle ground state of simple
Slater-determinant form. However, that the
numbers $I_j^{\alpha }$ of the RHS of Eq. $(8)$ can
be integers or half integers for different ground
states which, for example, differ by
$1$ in one of the eigenvalues of the diagonal
generators $(9)$, makes the problem much more
involved than a simple non-interacting case. This
change in the integer or half-integer character of
some of the numbers $I_j^{\alpha }$ of two
ground states, shifts {\it all} the occupied pseudomomenta
and leads to the orthogonal catastrophes \cite{Ander}.
This is the reason for the absence of quasiparticle peaks
in the single-particle spectral function. On the
other hand, if we consider the evaluation of quantities
which do not change the eigenvalues of the diagonal generators
$(9)$, the integer or half-integer character
of the numbers $I_j^{\alpha }$ is, in the thermodynamic
limit, irrelevant.

In some problems it is important to distinguish the
two situations $q_{F\alpha }^{(+)}=-q_{F\alpha }^{(-)}$
and $q_{F\alpha }^{(+)}\neq -q_{F\alpha }^{(-)}$.
For instance, the total-momentum expression $(22)$ below
reveals that when $q_{F\alpha }^{(+)}\neq -q_{F\alpha
}^{(-)}$ the $\alpha $ pseudoparticles give a finite
contribution to the momentum of the ground state $(11)$.
However, in many evaluations we can assume the
simplest case $q_{F\alpha }^{(+)}=-q_{F\alpha }^{(-)}$
and similarly for the pseudo-Brillouin zones limits.
This is because in either case
$q_{F\alpha }^{(+)}-q_{F\alpha }^{(-)}=2q_{F\alpha}-2\pi /N_a$
and $q_{\alpha }^{(+)}-q_{\alpha }^{(-)}=2q_{\alpha}-2\pi /N_a$,
and, in addition, except for terms of
order $1/N_a$, we have that $q_{F\alpha }^{(+)}=-q_{F\alpha }^{(-)}
\equiv q_{F\alpha }$ and $q_{\alpha }^{(+)}=
-q_{\alpha }^{(-)} \equiv q_{\alpha }$, where

\begin{equation}
q_{F\alpha } = {\pi N_{\alpha }\over {N_a}}
\, , \hspace{1cm} q_{\alpha } =
{\pi N_{\alpha }^*\over {N_a}} \, .
\end{equation}

The representation $(8)$ was used by Yang and Yang in
Ref. \cite{Yang69}. In the thermodynamic limit we take the
continuum limit $q_j\rightarrow q$ and the set of possible
pseudomomenta $(8)$ maps into a continuum domain of
$q$ values. For the calculation of many quantities
we can then replace $\pm q_{F\alpha }^{(\pm)}$
and $\pm q_{\alpha }^{(\pm)}$ by the values
$q_{F\alpha }$ and $q_{\alpha }$, respectively,
given in Eq. $(15)$. For instance, in Refs.
\cite{Carmelo92c,Carmelo91b,Carmelo92b}
that approximation was used in the evaluation of the
Landau-functional expansions because
this leads to the {\it correct} values for
the corresponding pseudoparticle bands and $f$ functions.
In contrast, however, when we insert in that functional particular forms of
the pseudoparticle deviations which refer to changes
in the numbers $N_{\alpha}$, the integer or half-integer
character of the quantum numbers $I_j^{\alpha }$ plays
a crucial role \cite{Frahm,Neto,Neto93}.

As in the case of Landau's Fermi liquid theory
\cite{Pines,Baym}, we will use the exact ground state $(11)$
of the quantum problem as reference state. In the
pseudoparticle basis we will write all operators in normal
order relative to that ground state. The corresponding
{\it universal form} of the normal-ordered quantum-liquid
Hamiltonian will confirm that in that basis the present
quantum system is a Landau liquid
\cite{Carmelo92,Carmelo92c,Carmelo91,Carmelo91b,Carmelo92b}.

It is useful to know how many LWS I and LWS II
there are for given numbers $\eta_z$ and $S_z$.
The total number of LWS, {\it i.e.} of regular BA Hamiltonian
eigenstates, was evaluated in Ref. \cite{Korepin}.
The following numbers refer both to canonical ensembles
of the present $U(1)\otimes U(1)$ sector and of the
$U(1)\otimes SU(2)$, $SU(2)\otimes U(1)$, and
$SO(4)$ sectors when $0\leq n\leq 1$ and
$0\leq m\leq n$. Only in the $U(1)\otimes U(1)$ sector
do all the LWS II states have an energy gap relative to the
ground state $(11)$. The number of LWS I and
LWS II is given by

\begin{equation}
\left[
\begin{array}{c}
N_a\\
N_s + N_s^*
\end{array}
\right]
\left[
\begin{array}{c}
N_s^*\\
N_s
\end{array}
\right] \, ,
\end{equation}
and

\begin{equation}
\left[
\begin{array}{c}
N_a\\
N_s^*
\end{array}
\right]\left[
\left[
\begin{array}{c}
N_a\\
N_s
\end{array}
\right]+
\left[
\begin{array}{c}
N_a\\
N_s - 2
\end{array}
\right]\right] - \left[
\left[
\begin{array}{c}
N_a\\
N_s^* + 1
\end{array}
\right] +
\left[
\begin{array}{c}
N_a\\
N_s^* - 1
\end{array}
\right]\right]
\left[
\begin{array}{c}
N_a\\
N_s - 1
\end{array}
\right] -
\left[
\begin{array}{c}
N_a\\
N_s + N_s^*
\end{array}
\right]
\left[
\begin{array}{c}
N_s^*\\
N_s
\end{array}
\right] \, ,
\end{equation}
respectively \cite{Two}. The square brackets in the above
equation refer to the usual combinatoric coefficients.
Adding the numbers $(16)$
and $(17)$, one recovers the total number of regular BA Hamiltonian
eigenstates for given values $\eta_z$ and $S_z$
\cite{Korepin}.

Since the colors $\alpha$ and the pseudomomentum $q$ are
the only quantum numbers involved in the description of the
pseudoparticles whose distributions define the
LWS I, these distributions can be
generated by applying to the vacuum $|V\rangle $ the
$\sum_{\alpha}N_{\alpha }$ creation operators $b^{\dag
}_{q\alpha}$. The resulting Hamiltonian eigenstates,
which are LWS I and of total number given by $(16)$,
include the ground state $(11)$ and the excited states
which can be generated from it by pseudoparticle-pseudoholes
processes and have the form

\begin{equation}
|\eta_z,S_z\rangle = \prod_{\alpha=c,s}
[\prod_{i,j=1}^{N_{ph}^{\alpha }}
b^{\dag }_{q_j\alpha }b_{q_i\alpha }]
|0;\eta_z,S_z\rangle \, ,
\end{equation}
where $q_j$ ($q_i$) defines the different locations of the
pseudoparticles (pseudoholes) relative to the reference
state $(11)$ and $N_{ph}^{\alpha }$ is the number of
$\alpha $ pseudoparticle-pseudohole processes.
$SO(4)$ multiplets are generated by acting the raising generators
$(4)$ and $(5)$ of the $\eta$ spin and spin algebras,
respectively, onto the LWS I
$(11)$ and $(18)$. As is shown in Ref. \cite{Nuno}, in
contrast to the higher-symmetry sectors, in the present
$U(1)\otimes U(1)$ sector we have that non-LWS multiplets
with values of $\eta_z$ and $S_z$ as in the corresponding
canonical ensemble have an energy gap relative to the
ground state $(11)$.

The LWS I states of the form $(11)$ and $(18)$ and of total
number given by
$(16)$ constitute a complete orthonormal basis which spans
an important Hilbert subspace, which we call $\cal {H}_I$.
At energy scales smaller than the gaps
for non-LWS multiplets \cite{Nuno} and LWS II, $\cal {H}_I$
represents the full accessible Hilbert space. We emphasize that
the generators which transform the ground state $(11)$ into
the excited eigenstates $(18)$ are products of {\it
one-pseudoparticle} operators of the form $b^{\dag }_{q+k\alpha}
b_{q\alpha}$, where $0<|q|<q_{F\alpha}^{(\pm)}$ and
$q_{F\alpha}^{(\pm)}<|q+k|<q_{\alpha}^{(\pm)}$.

Equations $(11)$ and $(18)$ reveal that in the basis associated
with the pseudoparticle operators $b^{\dag }_{q\alpha}$ and
$b_{q\alpha}$, all Hamiltonian eigenstates of the
many-electron system which are LWS I have a ``non-interacting'' form.
Since for a given canonical ensemble of values $(\eta_z,S_z)$ all these
excitations can be generated by successive applications of the operator
$b^{\dag }_{q+k\alpha}b_{q\alpha}$ to the ground
state $(11)$, the expression for this operator in the
usual electronic basis would provide interesting information
about the nature of the states $(18)$ from the point of view of
electronic configurations: while at large
$\pm 2q_{F\alpha}$ momenta and low energy the one-pair $\alpha$
pseudoparticle-pseudohole Hamiltonian eigenstates can correspond
to multipair electronic excitations \cite{Campbell,Carmelo94},
at small values of the momentum the operators
$b^{\dag }_{q+k\alpha}b_{q\alpha}$ are a superposition
of the two one-pair electronic (or particle)
operators $(4)$ \cite{Two,Carmelo94}. Fortunately, the
BA solutions are naturally expressed in terms of
the pseudoparticle basis.
Therefore, although we do not know, in general, the precise
form of the above generators in the electronic representation, these
solutions provide the expression for the Hamiltonian and other operators
in the pseudoparticle basis, and we can extract relevant
information about the quantum system without describing the
problem in terms of electronic configurations.
An important limitation, however, is that the BA solution
provides the pseudoparticle expressions for {\it some}
operators only. The general problem of the electron -
pseudoparticle canonical transformation will be studied
elsewhere \cite{Two,Carmelo94}. The fact that the BA solutions
are not most naturally expressed in terms of
the original electronic basis is the main reason why it has been
difficult, in previous work, to extract the information concerning
correlation functions and matrix elements contained
in the BA solutions.

Despite the non-interacting form of the Hamiltonian
eigenstates $(11)$ and $(18)$, we find in Sec. IV that the
normal-ordered Hamiltonian includes
pseudoparticle interaction terms and is, therefore,
a many-pseudoparticle operator. However, these pseudoparticle
interactions have a pure forward-scattering, zero-momentum transfer,
character. This agrees with the Landau-liquid studies of Refs.
\cite{Carmelo92,Carmelo92c,Carmelo91b,Carmelo92b},
which established this result using eigenenergies and eigenfunctions
rather than operators.

The expression of the Hamiltonian $(1)$ in $\cal {H}_I$
involves exclusively the two ($\alpha =c,s$)
pseudomomentum distribution operators

\begin{equation}
\hat{N}_{\alpha}(q)=
b^{\dag }_{q\alpha}b_{q\alpha} \, ,
\end{equation}
which play a key role in the present basis.
The operators $(19)$ commute with each other, i.e.
$[\hat{N}_{\alpha}(q),\hat{N}_{\alpha'}(q')]=0$. Note,
however, that the two (or $\nu$) pseudomomentum distribution
operators $(19)$ are {\it not} the two
momentum distribution electron (or particle) operators
$\hat{N}_{\sigma}(k)=c_{k\sigma}^{\dagger}c_{k\sigma}$.
The operator $\hat{N}_{\alpha}(q)$ ($\hat{N}_{\sigma}(k)$), which
has a simple form in the pseudoparticle (electronic) basis, has an
involved expression in the electronic (pseudoparticle)
basis \cite{Carmelo94}.

Since in the pseudoparticle basis the Hamiltonian expression
involves only the operators $(19)$, it follows that in
$\cal {H}_I$ the Hamiltonian {\it commutes}
with these operators. These play
a central role in this Hilbert subspace because all the
Hamiltonian eigenstates which are LWS I are also eigenstates of
$\hat{N}_{\alpha}(q)$. These LWS I are of form $(11)$ or
$(18)$ and, therefore, obey eigenvalue equations of the form

\begin{equation}
\hat{N}_{\alpha}(q)|\eta_z,S_z\rangle =
N_{\alpha}(q)|\eta_z,S_z\rangle \, ,
\end{equation}
where $N_{\alpha}(q)$ represents the real
eigenvalues of the operators $(19)$, which are given by
$1$ and $0$ for occupied and empty values of $q$, respectively.

It follows from Eq. $(9)$ that in $\cal {H}_I$
and in the pseudoparticle basis the diagonal generators $(2)$
of the two $U(1)$ diagonal subalgebras become

\begin{equation}
\hat{\eta}_z = -{1\over 2}
\sum_{q}[1 - \hat{N}_c(q)] \, , \hspace{1cm}
\hat{S}_z = - {1\over 2}\sum_{q}[1 -
\hat{N}_s(q)] \, .
\end{equation}

When we take the zero limit in one or several of the
eigenvalues of the diagonal generators, i.e. limit
$\eta_z\rightarrow 0$ or (and) $S_z\rightarrow 0$, we
approach a boundary line sector of parameter
space of higher symmetry. Following Eqs. $(9)$
and $(21)$, this corresponds to taking one or several of the
limits $N_{\alpha}\rightarrow N_{\alpha}^*$. If, for example,
we take this limit for one of the $\alpha$
pseudoparticle branches only, since the corresponding
number of $\alpha$ pseudoholes vanishes, Eqs. $(15)$ and $(16)$
reveal that the number of $\alpha$ configurations
reduces to $1$. This is a filled sea of $\alpha$ pseudoparticles
similar to that of the corresponding ground state (see Eqs. $(11)$ and
$(15)$). In particular, when we take the limit
$N_{\alpha}\rightarrow N_{\alpha}^*$ for all $\alpha$ pseudoparticle
branches, Eq.$(16)$ gives one single configuration which
is the $SO(4)$ half-filling and zero-magnetic field
ground state \cite{Two}. In this case there are no
excited LWS I $(18)$ and, except for the ground
state, {\it all} LWS
are LWS II \cite{Essler,Two}. On the other hand, it
also follows from Eq. $(16)$ that taking the limit
$N_{\alpha}\rightarrow 0$ in all pseudoparticle branches
corresponds also to a single configuration. In this case all
$\alpha$ bands are empty, i.e. full of pseudoholes. This
single state is the vacuum $|V\rangle$, which is both
the electronic and pseudoparticle vacuum.

These arguments show that, when we reach a parameter-space sector of
higher symmetry and/or a new phase by changing the
eigenvalue of one of the diagonal generators $(21)$, the
corresponding $\alpha$ branch of pseudoparticle-pseudohole
eigenstates {\it disappears}. In the case of a phase
transition, the collapsing branch is {\it not} replaced by a
new gapless branch of LWS II, whereas in the ``microscopic
transitions'' \cite{Carmelo92c,Two} one of the branches of LWS II becomes
gapless. In the latter case the number
of gapless branches remains the same, but the symmetry is
higher.  In the sector of parameter space of lowest symmetry
we have that the number of $\alpha$ configurations in $(16)$
is larger than one for all colors $\alpha$. Moreover, all
non-LWS multiplets \cite{Nuno} and LWS II, the study
of which we have omitted from this paper, have an energy gap
\cite{Carmelo92c,Carmelo91b,Two,Nuno94}.

Equation $(21)$ gives the expression of the diagonal
generators $(2)$ in the pseudoparticle basis.
Combining the operator representation that we
have just introduced with the information
contained in the BA solution, it is
straightforward to write other simple operators in
the pseudoparticle basis. The momentum $\hat{P}$ and
and the number of $\alpha $ pseudoparticle operator
$\hat{N}_{\alpha}$, for example, read

\begin{equation}
\hat{P} = \sum_{q,\alpha} q \hat{N}_{\alpha}(q) \, ,
\hspace{1cm}
\hat{N}_{\alpha} = \sum_{q} \hat{N}_{\alpha}(q) \, ,
\end{equation}
respectively. These expressions were derived by combining
Eqs. $(10)$ and $(20)$ with Eqs. $(8)-(10)$ of
Ref. \cite{Carmelo91b}. Note that these operators contain
non-interacting pseudoparticle terms only.

To close this section, and in order to clarify the relation of
our operator representation to the Landau-liquid functional of Refs.
\cite{Carmelo92,Carmelo92c,Carmelo91b,Carmelo92b},
we discuss the eigenvalues $N_{\alpha }(q)$ of Eq. $(20)$. From
Eq. $(22)$ we observe that these eigenvalues
obey the following normalization equation

\begin{equation}
N_{\alpha } = \sum_{q} N_{\alpha }(q) =
{N_a\over {2\pi}}\int_{q_{\alpha}^{(-)}}^{q_{\alpha}^{(+)}}dq
N_{\alpha }(q) \, .
\end{equation}
For instance, in the case of the ground state $(11)$ we have
that

\begin{eqnarray}
N_{\alpha}^0(q) & = & \Theta (q_{F\alpha}^{(+)}-q) \, ,
\hspace{0.5cm} 0<q<q_{\alpha}^{(+)}\nonumber \\
& = & \Theta (q-q_{F\alpha}^{(-)}) \, ,
\hspace{0.5cm} q_{\alpha}^{(-)}<q<0 \, ,
\end{eqnarray}
where $q_{F\alpha}^{(\pm)}$ is defined in Eqs.
$(12)-(14)$ and $q_{\alpha}^{(\pm)}$ is given by
similar equations with $N_{\alpha}$ replaced by
$N_{\alpha}^*$.

Using the ground state $(11)$ as the reference state,
among all the excited eigenstates of the form $(18)$ we
will be mostly interested in low-energy excitations. These
involve a redistribution of a {\it small density} of
pseudoparticles relative to the distribution $(24)$. Let
us introduce the normal-ordered pseudomomentum distribution
operator

\begin{equation}
:\hat{N}_{\alpha}(q): =
\hat{N}_{\alpha}(q) - N_{\alpha}^0(q) \, ,
\end{equation}
which is such that $\langle 0;\eta_z,S_z|:\hat{N}_{\alpha}(q):
|0;\eta_z,S_z\rangle=0$. The normal-ordered operators $(25)$
obey the eigenvalue equations

\begin{equation}
:\hat{N}_{\alpha}(q):|\eta_z,S_z\rangle =
\delta N_{\alpha}(q)|\eta_z,S_z\rangle \, ,
\end{equation}
where $|\eta_z,S_z\rangle$ denotes any Hamiltonian eigenstate of
form $(11)$ or $(18)$. Here $\delta N_{\alpha}(q)=N_{\alpha}(q)-
N_{\alpha}^0(q)$ and, therefore, $\delta N_{\alpha}(q)=0$ when
$|\eta_z,S_z\rangle=|0;\eta_z,S_z\rangle$.
The eigenvalues $N_{\alpha}(q)$ and $\delta
N_{\alpha}(q)$ are nothing but the pseudomomentum
distributions and deviations, respectively, of the Landau-liquid
theory studied in Refs.
\cite{Carmelo92,Carmelo92c,Carmelo91,Carmelo91b,Carmelo92b}.
Equations $(25)$ and $(26)$ imply that these are the
expectation values:

\begin{equation}
N_{\alpha}(q) = \langle \eta_z,S_z|\hat{N}_{\alpha}(q)
|\eta_z,S_z\rangle
\, ,
\end{equation}
and

\begin{equation}
\delta N_{\alpha}(q) = \langle \eta_z,S_z|:\hat{N}_{\alpha}(q):
|\eta_z,S_z\rangle
\, ,
\end{equation}
respectively. Therefore, the introduction of the operator
algebra $(7),(11)$, and $(18)$ clarifies the deep reasons for the
validity of the one-dimensional Landau-liquid theory, as
discussed in more detail in Sec. IV.

\section{RAPIDITY OPERATORS}

In this section we continue to restrict our study to $\cal {H}_I$.
As discussed above, in the case of LWS I,  the BA solutions of the
Hubbard model lead to two systems of algebraic equations.
Given the configuration of quantum numbers
that describes each Hamiltonian eigenstate
of form $(11)$ or $(18)$ (LWS I), these equations fully define the two
sets of rapidities which determine the corresponding
spatial wave function. As in the case of the quantum numbers
$I_j^{\alpha}$, each of these types of rapidities
is associated with one of the colors $\alpha$.
In the thermodynamic limit, we can take the pseudomomentum continuum
limit $q_j\rightarrow q$ and the rapidities
become functions of $q$ which we call $R_{\alpha} (q)$.
If we combine the operator representation introduced in
Sec. II with the properties of the BA solution, it is
straightforward to show
from the relation between the rapidities
$R_{\alpha} (q)$ and the pseudomomentum distributions $(19)$
and $(25)$ that these rapidity functions are nothing but real
{\it eigenvalues} of {\it rapidity operators}
$\hat{R}_{\alpha}(q)$ such that

\begin{equation}
\hat{R}_{\alpha}(q)|\eta_z,S_z\rangle =
R_{\alpha} (q)|\eta_z,S_z\rangle \, .
\end{equation}
The two rapidity operators $\hat{R}_{\alpha}(q)$
contain {\it all} information about the many-pseudoparticle
interactions of the quantum-liquid Hamiltonian. There are two
fundamental properties which imply the central role that the
rapidity operators of Eq. $(29)$ have in the present
quantum problem:

(a) Each of the normal-ordered rapidity operator $:\hat{R}_{\alpha}(q):$
can be written exclusively in terms of the two
pseudomomentum distribution operators $(25)$;

(b) The normal-ordered Hamiltonian can be written, exclusively,
in terms of the two pseudomomentum
distribution operators $(25)$, but all the corresponding
many-pseudoparticle interaction terms can be written in terms
of the rapidity operators $:\hat{R}_{\alpha}(q): $.
It follows that in the $\cal {H}_I$  the rapidity operators
commute with the Hamiltonian.

In the thermodynamic limit the BA algebraic equations
for the LWS I are replaced by a system of two
coupled integral equations
\cite{Yang,Lieb}. The standard treatment of the BA
solutions is very lengthy, {\it e.g.}, for each eigenstate we have to
{\it rewrite} a new set of equations. This is because the
eigenvalues $R_{\alpha} (q)$ of the rapidity
operators $\hat{R}_{\alpha}(q)$ are different for each state
$|\eta_z,S_z\rangle$ $(18)$.

However, within the present operator description,
we can introduce a single set of two
general operator equations which apply to {\it all} eigenstates.
Each of these equations defines one of the rapidity
operators $\hat{R}_{\alpha}(q)$ in terms of a $q$ and $\alpha$
summation containing functionals of the rapidity operators and
pseudomomentum-distribution operators. This system
of coupled equations has a unique solution which
defines the expressions for the rapidity operators in
terms of the pseudomomentum-distribution
operators. Although the structure of these equations leads to
many universal features which we will discuss later, they involve
the {\it spectral parameters} which are {\it not} universal but are
specific to each model. Also the formal expression for the Hamiltonian
in terms of the rapidity operators is not
universal. However, the corresponding normal-ordered expressions
do have a universal form for the integrable
multicomponent systems, as we shall discuss in Sec. IV.

For consistency with the previous notation
\cite{Carmelo92c,Carmelo91b,Carmelo92b}, we use
for the Hubbard chain the notation
$\hat{R}_c(q)=\hat{K}(q)$ and $\hat{R}_s(q)=\hat{S}(q)$. The
spectral parameters are the numbers $(4t/U)\sin [K(q)]$
and $S(q)$, which appear in Appendix A and in Eqs.
$(30)-(32)$ below in operator form.

We have mentioned several times that the BA solution
is most naturally expressed in the pseudoparticle basis.
One reflection of this is the simple expression for
the Hamiltonian $(1)$ in that basis. Combining Eq. $(29)$
with the energy expression $(4)$ of Ref. \cite{Carmelo92b},
which refers to the LWS I, the result is

\begin{equation}
\hat{H} = \sum_{q}\hat{N}_c(q)\{-2t\cos [\hat{K}(q)]
- U/2\} - 2\mu\hat{\eta }_z - 2\mu_0 H\hat{S}_z \, ,
\end{equation}
where the expressions of the diagonal generators are given in
Eq. $(21)$. This is the exact expression of the Hamiltonian $(1)$
in $\cal {H}_I$. At energy scales smaller than the gaps for the
non-LWS multiplets and LWS II, $(30)$ gives the exact expression
of that Hamiltonian in the full Hilbert space.
Despite its simple appearance, the Hamiltonian $(30)$
describes a many-pseudoparticle problem. The reason is that the
expression of the rapidity operator $\hat{K}(q)$ in terms of
the operators $\hat{N}_{\alpha}(q)$ contains many-pseudoparticle
interacting terms.

To specify the Hamiltonian $(30)$ completely, we must indicate
the operator equations that define the rapidity operators
in terms of the pseudoparticle momentum distribution operators.
In the case of the Hubbard chain these two operator equations
read

\begin{equation}
[\hat{K}(q) - {2\over N_a}\sum_{q'}\hat{N}_s(q')
\tan^{-1}\Bigl(\hat{S}(q') - (4t/U)\sin
[\hat{K}(q)]\Bigr)]|\eta_z,S_z\rangle = q|\eta_z,S_z\rangle
\end{equation}
and

\begin{eqnarray}
&  & {2\over N_a}[\sum_{q'}\hat{N}_c(q')
\tan^{-1}\Bigl(\hat{S}(q) - (4t/U)
\sin [\hat{K}(q')]\Bigr)
\nonumber \\
& - & \sum_{q'}\hat{N}_s(q')\tan^{-1}\Bigl({1\over 2}
\left(\hat{S}(q) - \hat{S}(q')\right)\Bigr)]|\eta_z,S_z\rangle =
q|\eta_z,S_z\rangle \, .
\end{eqnarray}
These equations fully define the rapidity operators in terms
of the pseudomomentum distribution operators $(19)$ and
$(25)$.

Simple generalizations of these equations hold
for other multicomponent quantum liquids:
the many-pseudoparticle terms of the Hamiltonian
can be expressed, exclusively, in terms of the rapidity
operators. Moreover, the BA solutions always provide:
(a) the expressions of the Hamiltonian in
terms of the rapidity operators; and (b)
$\nu$ operator equations which define (implicitly) the expressions
of the rapidity operators in terms of the pseudomomentum
distribution operators $(19)$.

The forms of the eigenfunction equations $(20)$ and $(29)$
show that the operator equations $(31)$ and $(32)$
are equivalent to the BA equations presented
previously \cite{Carmelo92c,Carmelo91b,Carmelo92b}.
Formally, these are obtained from Eqs. $(31)$ and $(32)$ by replacing
the operators by the corresponding eigenvalues and projecting
onto the various $|\eta_z,S_z\rangle$ states. As in the case
of Eqs. $(31)$ and
$(32)$, these equations are general and apply to all
eigenstates. Their solution gives the rapidity-Landau
functionals in terms of the pseudomomentum distributions
$(19)$. Insertion into these functionals of the pseudomomentum
distributions of a given Hamiltonian eigenvalue $|\eta_z,S_z\rangle$
leads to the corresponding rapidity eigenvalue of the
RHS of Eq. $(29)$.

Unsurprisingly, it is difficult to solve the BA operator equations
$(31)-(32)$ directly and to obtain the explicit expression for the rapidity
operators in terms of the pseudomomentum distribution operators
$(19)$. In contrast, it is easier to calculate their normal-ordered
expression in terms of the normal-ordered operators $(25)$. In
the ensuing section, we introduce the pseudoparticle perturbation theory
which leads to the normal-ordered expressions for the
Hamiltonian and rapidity operators.

\section{PSEUDOPARTICLE PERTURBATION THEORY}

In the pseudoparticle basis the normal-ordered rapidity
operators $:\hat{R}_{\alpha}(q):$ contain
an infinite number of terms, as we shall demonstrate below. The first of
these terms is linear in the pseudomomentum distribution
operator $:\hat{N}_{\alpha}(q):$ $(25)$, whereas the remaining terms
consist of products of two, three,....., until infinity,
of these operators. The number of $:\hat{N}_{\alpha}(q):$ operators
which appears in these products equals the order of the
scattering in the corresponding rapidity term.

A remarkable property is that in the pseudoparticle
basis the seemingly ``non-perturbative'' quantum liquids
become {\it perturbative}: while excitations associated with
adding or removing of one electron have zero life-time and decay
into collective pseudoparticle excitations, all low-energy
Hamiltonian eigenstates of the system with one electron more or less
are pseudoparticle-pseudohole states of the
form $(11)$ or $(18)$. Furthermore, while the two-electron
forward scattering amplitudes and vertices diverge, the
two-pseudoparticle $f$ functions (given by Eq. $(42)$
below) and the corresponding two-pseudoparticle
forward-scattering amplitudes, which were calculated in
Ref. \cite{Carmelo92c}, are finite. By ``perturbative'' we also
mean here the
following: since at each point of parameter space (canonical
ensemble) the excited low-energy eigenstates are of form $(18)$
and correspond to quantum-number configurations involving a
density of excited pseudoparticles relative to the ground-state
configuration $(11)$, $(24)$, the expectation values of the Hamiltonian
in these states are functions of the density
of excited pseudoparticles. This density is given by

\begin{equation}
n_{ex} = \sum_{\alpha}n_{ex}^{\alpha} \, ,
\end{equation}
where

\begin{equation}
n_{ex}^{\alpha} = {1\over {N_a}}\sum_{q}
[1 - N_{\alpha}^0(q)]\delta N_{\alpha}(q)
\end{equation}
defines the density of excited $\alpha$ pseudoparticles (and
$\alpha$ pseudoholes)
associated with the Hamiltonian eigenstate $|\eta_z,S_z \rangle$.

When all the densities $n_{ex}^{\alpha}$ are small, we can
expand the expectation values in these densities. The
perturbative character of the quantum liquid rests on the
fact that the evaluation of the expectation values up to the
$ n^{th}$ order in the densities $(33)$ requires considering only the
corresponding operator terms of scattering orders less than or
equal to n. This follows from the linearity of the density of
excited $\alpha$ pseudoparticles, which are the elementary
``particles'' of the quantum liquid, in
$\delta N_{\alpha}(q)=\langle\eta_z,S_z|:\hat{N}_{\alpha}(q):
|\eta_z,S_z \rangle$
and from the form of $(34)$. The perturbative character of the quantum
liquid implies, for example, that, to second order in the density
of excited pseudoparticles, the energy involves only one-
and two-pseudoparticle Hamiltonian terms \cite{Neto,Neto93},
as in the case of the quasiparticle terms of a Fermi-liquid
energy functional \cite{Pines,Baym}.

We note parenthetically that Eqs. $(33)$ and $(34)$ do not apply
to the excitations
(A) studied in \cite{Neto} and in II. In the
case of these excitations, which are associated with changes in
the numbers $N_{\alpha}$, the densities $(33)$ and $(34)$ are
replaced by the density of ``removed'' or ``added'' pseudoparticles.
In this case the change in the integer or half-integer
character of the quantum numbers must be taken into
account.

The eigenvalue equations $(26)$ imply that the problem of using
the rapidity Eqs. $(31)$ and $(32)$ to derive the
expression of the operators $:\hat{R}_{\alpha}(q):$
in terms of the operators $:\hat{N}_{\alpha}(q):$ is equivalent
to the problem of evaluating the corresponding expansion
of the rapidity eigenvalues $\delta R_{\alpha}(q)$ in terms of the
pseudoparticle deviations $\delta N_{\alpha}(q)$ $(28)$. This last
problem, which leads to the Landau-liquid expansions, was studied
in Ref. \cite{Carmelo92b} for the case of the Hubbard chain.
Furthermore, we emphasize that it is the perturbative character of
the pseudoparticle basis which {\it justifies} the validity of
these Landau-liquid deviation expansions
\cite{Carmelo92c,Carmelo91,Carmelo91b,Carmelo92b}.

Based on the connection between the two problems we can
derive the expressions for the rapidity operators
$:\hat{R}_{\alpha}(q):$. This corresponds to expanding the
expressions of the operators $:\hat{R}_{\alpha}(q):$ in terms
of increasing pseudoparticle scattering order. It is convenient
to define these expressions through the operators
$:\hat{Q}_{\alpha}(q):$. These are related to the rapidity
operators as follows

\begin{equation}
:\hat{R}_{\alpha}(q): = R_{\alpha}^0(q + :\hat{Q}_{\alpha}(q):)
- R_{\alpha}^0(q) \, ,
\end{equation}
where $R_{\alpha}^0(q)$ is the ground-state eigenvalue
of $\hat{R}_{\alpha}(q)$, i.e.

\begin{equation}
\hat{R}_{\alpha}(q)|0;\eta_z,S_z\rangle =
R_{\alpha}^0(q)|0;\eta_z,S_z\rangle \, .
\end{equation}
The operators $:\hat{Q}_{\alpha}(q):$ contain the same
information as the rapidity operators, and involve, exclusively, the
two-pseudoparticle phase shifts $\Phi_{\alpha\alpha
'}(q,q ')$ defined in Ref. \cite{Carmelo92b}. Following
the related studies of Refs. \cite{Carmelo92c,Carmelo91b,Carmelo92b},
in Appendix A we introduce Eq. $(35)$ in the BA equations
$(31)$ and $(32)$ and expand in the scattering order. This
leads to

\begin{equation}
:\hat{Q}_{\alpha}(q): = \sum_{i=1}^{\infty}
\hat{Q}_{\alpha}^{(i)}(q) \, ,
\end{equation}
where $i$ gives the scattering order of the operator
term $\hat{Q}_{\alpha}^{(i)}(q)$. For example, for the
first-order term we find

\begin{equation}
\hat{Q}_{\alpha}^{(1)}(q) = {2\pi\over {N_a}}
\sum_{q',\alpha'}\Phi_{\alpha\alpha '}(q,q '):\hat{N}_{\alpha'}(q'):
\, .
\end{equation}
We emphasize that while the expressions for the phase shifts
$\Phi_{\alpha\alpha '}(q,q ')$ are specific to each model
because they involve the spectral parameters
\cite{Carmelo92b}, the {\it form} of the operator term
$\hat{Q}_{\alpha}^{(1)}(q)$ $(38)$ is {\it universal}
and refers to the multicomponent quantum liquids discussed in Secs. I and V.

All the remaining higher-order operator terms of expression
$(37)$, $\hat{Q}_{\alpha}^{(i)}(q)$, can be obtained from the
rapidity equations $(31)$ and $(32)$. For simplicity, we provide
only the expression of the first-order operators $(38)$ and
of the second-order operators, which are given in Eqs.
(A20)-(A23) of Appendix A. In that
Appendix we evaluate the expressions of these operators
and use the Hamiltonian expression $(30)$ in terms of the
rapidity operators to derive the following expression for
the normal-ordered Hamiltonian :

\begin{equation}
:\hat{H}: = \sum_{i=1}^{\infty}\hat{H}^{(i)} \, ,
\end{equation}
where, to second pseudoparticle scattering order

\begin{eqnarray}
\hat{H}^{(1)} & = & \sum_{q,\alpha}
\epsilon_{\alpha}(q):\hat{N}_{\alpha}(q): \, ;\nonumber\\
\hat{H}^{(2)} & = & {1\over {N_a}}\sum_{q,\alpha} \sum_{q',\alpha'}
{1\over 2}f_{\alpha\alpha'}(q,q')
:\hat{N}_{\alpha}(q)::\hat{N}_{\alpha'}(q'): \, .
\end{eqnarray}
Here $(40)$ are the Hamiltonian terms which are {\it
relevant} at low energy \cite{Neto,Neto93}. Furthermore,
it is shown in Refs. \cite{Neto,Neto93} that at
low energy and small momentum the only relevant term is the
non-interacting term $\hat{H}^{(1)}$. This property justifies to the
Landau-liquid character of these systems and plays a key role
in the symmetries of the critical point \cite{Neto,Neto93}.

The form of the normal-ordered Hamiltonian $(39)-(40)$ is
{\it universal} for the
integrable multicomponent quantum
liquids. On the other hand, the expressions for the pseudoparticle
bands $\epsilon_{\alpha}(q)$ involve the spectral parameters and
are specific to each model. For the case of the Hubbard chain
they are defined in Ref. \cite{Carmelo91b}. The $c$ and $s$
pseudoparticle bands are shown in Figs. 7 and 8,
respectively, of that reference.

All $\hat{Q}_{\alpha}^{(i)}(q)$ terms of the RHS of Eq. $(37)$
are such that both the $f$ functions of the RHS of Eq. $(40)$
and all the remaining higher order coefficients associated with the
operators $\hat{H}^{(i)}$ of order $i>1$ have {\it universal}
forms in terms of the {\it two-pseudoparticle} phase shifts and
pseudomomentum derivatives of the bands and coefficients
of order $<i$. This follows from the fact that the $S$-matrix for
$i$-pseudoparticle scattering factorizes into two-pseudoparticle
scattering matrices, as in the case of the usual BA
$S$-matrix \cite{Essler,Korepin79,Zamo,Andrei}. For
example, although the second-order term $\hat{H}^{(2)}$ of
Eq. $(40)$ involves an integral over the second-order
function $\hat{Q}_{\alpha}^{(2)}(q)$ (see Eq. (A24) of
Appendix A), this function is such that $\hat{H}^{(2)}$
can be written {\it exclusively} in terms of the
first-order functions $(38)$,

\begin{equation}
\hat{H}^{(2)} = \sum_{q,\alpha}v_{\alpha} (q)
\hat{Q}_{\alpha}^{(1)}(q) :\hat{N}_{\alpha}(q):
+ {N_a\over {2\pi}}\sum_{\alpha}{v_{\alpha}\over 2}
\sum_{j=\pm 1}[\hat{Q}_{\alpha}^{(1)}(jq_{F\alpha})]^2 \, ,
\end{equation}
as shown in Appendix A, and the four (or, in the general case,
$\nu\times\nu$) ``Landau'' $f$ functions, $f_{\alpha\alpha'}(q,q')$,
have universal forms which read

\begin{eqnarray}
f_{\alpha\alpha'}(q,q') & = & 2\pi v_{\alpha}(q)
\Phi_{\alpha\alpha'}(q,q')
+ 2\pi v_{\alpha'}(q') \Phi_{\alpha'\alpha}(q',q) \nonumber \\
& + & \sum_{j=\pm 1} \sum_{\alpha'' =c,s}
2\pi v_{\alpha''} \Phi_{\alpha''\alpha}(jq_{F\alpha''},q)
\Phi_{\alpha''\alpha'}(jq_{F\alpha''},q') \, ,
\end{eqnarray}
where the pseudoparticle group velocities are given by

\begin{equation}
v_{\alpha}(q) = {d\epsilon_{\alpha}(q) \over {dq}} \, ,
\end{equation}
and depend on the spectral parameters. In particular, the
velocities

\begin{equation}
v_{\alpha}\equiv v_{\alpha}(q_{F\alpha}) \, ,
\end{equation}
play a determining role at the critical point, representing the ``light''
velocities which appear in the conformal-invariant expressions
\cite{Frahm,Neto,Neto93}.
In the case of the Hubbard chain, the velocities $(44)$ are plotted
in Fig. 9 of Ref. \cite{Carmelo91b}.

We note that the Hamiltonian term $\hat{H}^{(1)}$ in
$(40)$ has, from the point of view of the pseudoparticle basis,
a non-interacting character. However, in the electron basis,
$\hat{H}^{(1)}$ is of interacting character, as revealed by the $U$
dependence of the bands $\epsilon_{\alpha}(q)$
\cite{Carmelo91b}. Furthermore, as we mentioned
above, both the two-pseudoparticle $f$ functions $(42)$
and forward-scattering amplitudes \cite{Carmelo92c} are
finite. This is in contrast to the non-perturbative electronic
basis, where the two-electron forward-scattering vertices
and amplitudes diverge.

Obviously, Eqs. $(27)$ and $(28)$ imply that the operator
expressions $(35)$ and $(37)-(40)$ are
fully equivalent to the corresponding Landau expansions
already studied in Refs. \cite{Carmelo92c,Carmelo91b,Carmelo92b}
for the case of the Hubbard chain. Thus we have achieved
one of the aims of the present paper, which is to use the
pseudoparticle operator algebra to justify
the validity of the Landau-liquid properties
of the quantum problem already studied in these
earlier papers.

We emphasize that in the present $U(1)\otimes U(1)$
sector of parameter space and at energy scales smaller than
the gaps for the non-LWS and LWS II,  Eqs. $(30)$ and $(39)-(40)$
refer to the expression of the full quantum-liquid Hamiltonian.
In the electronic basis this is given by Eq. $(1)$.
Our operator representation leads, in a natural way, to the
low-energy spectrum studied in Refs.
\cite{Frahm,Carmelo92,Carmelo92c,Carmelo91b,Carmelo92b}. In
particular, the study of Refs.
\cite{Carmelo92,Carmelo92c,Carmelo91b,Carmelo92b} shows
that the low-energy of the Hamiltonian $(39)-(40)$ is, in many
points, similar to that of a Fermi liquid, with the
pseudoparticles playing the role of the quasiparticles
and the colors $\alpha$ of the spin projections
$\sigma$. For example, the static
susceptibilities can be calculated as in a
Fermi liquid \cite{Carmelo92b}, and involve, exclusively, the
velocities $(44)$ and the expressions of
the two-pseudoparticle phase shifts at the pseudo-Fermi
points. Also the
low-temperature thermodynamics can be studied as in a Fermi
liquid, being described by pseudoparticle Fermi-Dirac
distributions of the form \cite{Carmelo91b}

\begin{equation}
N_{\alpha}(q)={1\over {1+e^{
\epsilon_{\alpha}(q)/k_BT}}} \, .
\end{equation}
As in a Fermi
liquid, the low-temperature specific heat is linear in
the temperature and involves the static masses
$m_{\alpha }^* = {q_{F\alpha } \over {v_{\alpha }}}$,
where the pseudo-Fermi pseudomomenta and velocities are defined
in Eqs. $(15)$ and $(44)$, respectively.
The low-frequency and low-momentum dynamical properties
can be studied by means of kinetic equations
\cite{Carmelo92,Carmelo92c}, again as in a Fermi liquid.
This allows the evaluation of the pseudoparticles elementary
currents and transport masses. The latter masses, the
expressions for which differ from the above expressions for the static
masses, define the low-energy {\it dynamical separation}
which characterizes the multicomponent quantum liquid
\cite{Carmelo92c,Neto,Neto93}. All this refers to
the coherent part of the conductivity. The incoherent
part is determined by the LWS II and non-LWS multiplets
\cite{Carmelo92c,Nuno94,Horsch94}.

In this section we have presented the expression of the
normal-ordered Hamiltonian and rapidity operators in the
pseudoparticle basis of $\cal {H_I}$. This confirms
the consistency of the one-dimensional Landau-liquid
theory which was shown to refer to this basis. The advantage
of using the pseudoparticle basis is that the problem
becomes perturbative, {\it i.e.} from the point of view of the
pseudoparticle interactions it is possible to classify
which scatterings are {\it relevant}. Consistent with
earlier results \cite{Neto} and our more detailed study in paper II,
while at low energy and {\it small} momentum {\it all}
pseudoparticle interactions are irrelevant, at low energy
and {\it large} momentum only the two-pseudoparticle interactions
are relevant. The same happens for low-energy excitations
involving changes in the numbers $N_{\alpha}$. For
simplicity and for further use in paper II,
we have presented in this paper only the two
first Hamiltonian terms of expression $(39)$. However, Eqs.
$(30)-(32)$ contain {\it full} information about
all the remaining terms of higher-scattering order.

In the companion paper II (see also \cite{Neto}), the perturbative
character of the present class of quantum liquids is used to
study the symmetries and algebras at the critical point.
These studies confirm that the Hamiltonian $(39)$ is
the {\it correct} starting point to construct a
critical-point Hamiltonian.

\section{CONCLUDING REMARKS}

In the previous sections we have introduced and studied
a new operator algebra \cite{Neto} describing the
low-energy Hamiltonian eigenstates of
integrable quantum liquids. Considering explicitly
the case of the Hubbard chain in a
magnetic field and chemical potential, we have
shown that the new operators
create and annihilate the pseudoparticles of the
one-dimensional Landau liquid \cite{Carmelo92,Carmelo92c}.
Our algebraic approach permits an operator analysis of the BA
solutions. Further, it allows us to
construct a normal-ordered Hamiltonian which has in the
pseudoparticle basis a universal form involving
only $k = 0$, forward-scattering pseudoparticle interaction terms.
Further, the perturbative character of the pseudoparticle
basis implies that the corresponding two-pseudoparticle Landau
$f$ functions and forward-scattering amplitudes are
finite, in contrast to the non-perturbative
electronic representation, in which the two-electron
forward-scattering amplitudes and vertices diverge.
To clarify further the context and implications of
our results, we close with several remarks.

First, concerning the relation to previous
work on pseudoparticles
\cite{Carmelo92,Carmelo92c,Carmelo91,Carmelo91b,Carmelo92b},
we note that the pseudoparticle operator algebra motivates and
justifies the validity of the Landau-liquid studies. These
studies were based on explicit use of eigenenergies and
eigenfunctions, rather than the general algebraic approach.

Second, although our explicit calculations were presented
for the Hubbard model, which in the general terminology corresponds
to a ``two-component integrable quantum fluid'', our results
in fact apply to the class of BA solvable models
\cite{Izergin,Frahm,Neto}, which in the largest sector of the
parameter space have symmetry $[U(1)]^{\nu}$ \cite{Neto}.
As remarked previously, in this sector, the model has
$\nu$ independent branches of
gapless elementary excitations. Therefore, when $\nu>1$
these models are usually referred as {\it multicomponent}
integrable quantum liquids \cite{Izergin,Frahm,Neto}.
Note in this sector that $\nu$ represents the
number of independent conserved quantum numbers
and also the number of
external fields (magnetic field, chemical potential, etc.)
which are ``conjugate'' to these conserved quantum numbers
in the statistical mechanics sense. Although some of these systems
are described by continuum models \cite{Yang}, most of them
are defined in an one-dimensional chain with $N_a$
sites, $j=1,...,N_a$, refer to interacting fermionic
particles, and, in particular, to interacting electrons,
as in the case of the Hubbard model studied in this paper.

More precisely, the class of solvable quantum liquids
to which our results apply is defined by the
fact that the $\nu\times\nu$ dressed-charge matrix
\cite{Izergin,Frahm,Carmelo92c,Neto} is determined
exclusively by the $R$ matrix associated with the solution of
the Yang-Baxter equation \cite{Izergin,Neto}.
In this regard, we note that our study here, when
supplemented by the results in II and Ref. \cite{Neto},
shows that the present operator algebra is
consistent with the finite-size-correction results
of Frahm and Korepin \cite{Frahm}, who found that the complete critical
theory of $\nu$ multicomponent integrable systems is a direct
product of $\nu$ Virasoro algebras.

Third, the pseudoparticle algebra allows us to see immediately
the origin of the ``universal'' character
of these one-dimensional integrable quantum liquids,
which can be understood as a straightforward
generalization to pseudoparticles of Wilson's
renormalization group arguments \cite{Castro}: the pseudo-Fermi points
of the pseudoparticles replace the particle Fermi points, and
close to the pseudo-Fermi points only few types of
two-pseudoparticle scattering processes are relevant
for the low-energy physics.

Fourth, concerning the very important problem of the
relation of pseudoparticles to the original electrons, we
should make several remarks. As we have stressed throughout,
the essential simplifying feature of the pseudoparticle basis
is that the quantum problem becomes perturbative there,
in the sense that it is possible to identify which
pseudoparticle scatterings are relevant for the low-energy
physics. For instance, in the new basis the two-pseudoparticle
forward-scattering vertices and amplitudes do not
diverge and one can define a many-pseudoparticle
perturbation theory in which the non-interacting ground state
is the exact ground state of the many-electron
problem. This is directly analogous to Landau's Fermi-liquid theory,
in which the interacting ground state is used as reference state
\cite{Pines,Baym}. As is well known, the
present class of quantum liquids are {\it not} Fermi liquids because
the single-particle spectral function has no quasiparticle
peak ({\it i.e.}, $Z_F = 0$), but they do have what
was previously called a `` Landau-liquid'' character
\cite{Carmelo92,Carmelo92c,Carmelo91,Carmelo91b,Carmelo92b}:
in the pseudoparticle basis the low-energy physics is fully
controlled by two-pseudoparticle forward scattering, and writing
the Hamiltonian and other operators in the pseudoparticle basis allows
the use of perturbation-theory techniques, as we have shown in this
paper and will exploit in II.

The non-perturbative character of the usual electronic basis --
{\it e.g.},  the divergences of the two-electron forward-scattering
vertices and amplitudes -- is reflected in the complex and
exotic properties of the electron-pseudoparticle
operator transformation, which we have studiously
avoided in this paper. The construction of this
energy, momentum, and parameter dependent nonlinear
transformation is in general very involved. For instance, we
know that at low energies and
large momenta it maps one-pair electron operators
onto multipair pseudoparticle operators, and vice versa
\cite{Campbell}.
Further, both the colors $\alpha $ and pseudomomentum
$q$ of the pseudoparticles are {\it not}, for
general values of the parameters, simply related to the
usual electronic quantum
numbers ({\it i.e.}, charge, spin, up spin, down spin,
and momentum) \cite{Two,Carmelo94}.
We shall return to this problem elsewhere
\cite{Carmelo94} and solve it for the generators
which transform the exact ground state onto the low-energy
and small momentum Hamiltonian eigenstates, which we are
able to write both in terms of pseudoparticle and electron
operators. Importantly, as we have demonstrated in our
calculations here, since the BA solution
refers at low energy to the pseudoparticle operator
representation, we can study the quantum
problem in the corresponding basis {\it without} explicit
knowledge of the complex electron-pseudoparticle transformation.

Fifth, concerning applications of our ideas to other
theoretical models and to problems in real materials,
we note that although our present analysis refers to a class of
one-dimensional integrable quantum liquids only, it
has been argued that similar non-Fermi liquid behavior occurs
in some sectors of parameter space of two-dimensional
interacting quantum liquids
\cite{Ander,Varma,Guinea,Essler92,Hua}. If this is so,
techniques similar to ours may apply. Further, it is of
considerable interest to determine the extent to which
the algebraic structure and pseudoparticle perturbation
theory remain valid in systems which are {\it not} integrable
but which behave as Luttinger (or Landau) liquids
(about the relation between Luttinger and Landau
liquids see Sec. VIII of Ref. \cite{Carmelo92c}); an example
of such a system is the one-dimensional extended Hubbard model.
In terms of real materials, it is
known that the present one-dimensional quantum
systems provide useful (albeit idealized) descriptions of the physics of
quasi-one-dimensional solids \cite{Campbell,Pouget,Jacob}.
At low frequency the pseudoparticles are the transport
carriers of the quantum liquid and couple to external fields
\cite{Carmelo92,Carmelo92c}. These couplings determine
the exotic instabilities observed in quasi-one-dimensional
synthetic metals \cite{Campbell}.

Finally, we turn to directions for further research. In
addition some specific calculations mentioned above, our
immediate aims are to apply the concepts and techniques
developed here to establish that the
pseudoparticle perturbation theory leads to the correct
effective Hamiltonian, which is obtained from the {\it universal}
expression $(39)-(40)$ of the quantum-liquid normal-ordered
Hamiltonian in the {\it pseudoparticle basis}, and to study the
Virasoro algebras \cite{Frahm,Neto} of conformal field theory.
The initial results of these studies are described in II.
The construction of a generalized Landau-liquid theory
referring to the Hilbert space spanned by both the LWS I
and LWS II is in progress.

\nonum
\section{ACKNOWLEDGMENTS}

This work was supported principally by the C.S.I.C. [Spain] and
the University of Illinois. We thank F. H. L. Essler, E. H. Fradkin,
F. Guinea,
P. Horsch, V. E. Korepin, A. A. Ovchinnikov, N. M. R. Peres, and
K. Maki for stimulating discussions. The hospitality and support
of the C.S.I.C. and U.I.U.C. are gratefully acknowledged by JMPC.
AHCN thanks CNPq (Brazil) for a scholarship.
DKC and AHCN acknowledge the partial support of this research
by the U.S. National Science Foundation under grants
NSF-DMR89-20538 and NSF-DMR91-22385, respectively.

\vfill
\eject
\appendix{NORMAL-ORDERED OPERATOR EXPRESSIONS}

Following the discussion of Sec. IV, the perturbative character of
the system implies the equivalence between expanding in the
pseudoparticle scattering order and/or in the pseudomomentum
deviations $(28)$. This, together with Eqs. $(20)$ and $(26)$,
reveals that the problem of calculating the normal-ordered operator
expansions $(37)$ and $(39)$ is equivalent to the Landau-energy
functional studies of Refs. \cite{Carmelo92c,Carmelo91b,Carmelo92b}.
Since in these papers the full expression of the
second-order function $Q_{\alpha}^{(2)}(q)$ (which corresponds
to the operator $\hat{Q}_{\alpha}^{(2)}(q)$) and a number of other
functions were not presented, we give in this Appendix a short
description of the calculation of the normal-ordered operator
expansions $(37)$ and $(39)$. We focus our brief study on
the points which were omitted in Refs.
\cite{Carmelo92c,Carmelo91b,Carmelo92b}.

We start by the evaluation of the first-order and second-order
terms of the operator $(37)$. For simplicity, we consider here
pseudomomentum deviations and eigenvalues. Equation $(26)$ then
allows the straightforward calculation of the corresponding
operator expressions.

In the thermodynamic limit, Eqs. $(31)$ and $(32)$ lead to
the following equations

\begin{equation}
K(q) - {1\over \pi}\int_{q_{Fs}^{(-)}}^{q_{Fs}^{(+)}}N_s(q')
\tan^{-1}\Bigl(S(q') - {\sin K(q)\over u}\Bigr) = q
\end{equation}
and

\begin{eqnarray}
&  & {1\over \pi}[\int_{q_{Fc}^{(-)}}^{q_{Fc}^{(+)}}N_c(q')
\tan^{-1}\Bigl(S(q) - {\sin K(q')\over u}\Bigr)
\nonumber \\
& - & \int_{q_{Fs}^{(-)}}^{q_{Fs}^{(+)}}
N_s(q')\tan^{-1}\Bigl({1\over 2}
\left(S(q) - S(q')\right)\Bigr)] = q \, ,
\end{eqnarray}
where we have replaced the operators by the corresponding
eigenvalues and the summations by integrations and $u=U/4t$.
The eigenvalue form of Eq. $(35)$ is

\begin{equation}
\delta R_{\alpha}(q) = R_{\alpha}^0(q + \delta Q_{\alpha}(q))
- R_{\alpha}^0(q) \, ,
\end{equation}
where $\delta R_{\alpha}(q)$ and $\delta Q_{\alpha}(q)$ are the
eigenvalues of the operators $:\hat{R}_{\alpha}(q):$ and
$:\hat{Q}_{\alpha}(q):$, respectively. From Eq. $(37)$
$\delta Q_{\alpha}(q)$ can be written as

\begin{equation}
\delta Q_{\alpha}(q) = Q_{\alpha}^{(1)}(q) + Q_{\alpha}^{(2)}(q)
+ ... \, ,
\end{equation}
where $Q_{\alpha}^{(i)}(q)$ is the eigenvalue of the operator
$\hat{Q}_{\alpha}^{(i)}(q)$. Expanding the $\delta R_{\alpha}(q)$
expression (A3) we find

\begin{equation}
R_{\alpha}(q) =  \sum_{i=0}^{\infty} R_{\alpha}^{(i)}(q) \, ,
\end{equation}
(and $\delta R_{\alpha}(q)=\sum_{i=1}^{\infty}
R_{\alpha}^{(i)}(q)$) where the zero-order ground-state
functions $R_{\alpha}^{(0)}(q)$ are defined by Eqs. (A1)
and (A2) with the distributions $N_{\alpha}(q)$
given by Eq. $(24)$. From the resulting equations we can
obtain all derivatives of the ground-state functions
$R_{\alpha}^{(0)}(q)$ with respect to $q$. The terms of the
RHS of Eq. (A5) involve these derivatives. For instance,
the first-order and second-order terms read

\begin{equation}
R_{\alpha}^{(1)}(q) =  {d R_{\alpha}^{(0)}(q)\over {dq}}
Q_{\alpha}^{(1)}(q) \, ,
\end{equation}
and

\begin{equation}
R_{\alpha}^{(2)}(q) =  {d R_{\alpha}^{(0)}(q)\over {dq}}
Q_{\alpha}^{(2)}(q) + {1\over 2}{d^2 R_{\alpha}^{(0)}(q)\over
{dq^2}}[Q_{\alpha}^{(1)}(q)]^2 \, ,
\end{equation}
respectively, and involve the first and second derivatives.
Let us use the notation of Eqs. (A1) and (A2), i.e.,

\begin{equation}
R_{c}(q) = K(q) \hspace{2cm} R_{s}(q) = S(q) \, ,
\end{equation}
and

\begin{equation}
R_{c}^{(i)}(q) = K^{(i)}(q) \hspace{2cm}
R_{s}^{(i)}(q) = S^{(i)}(q) \, .
\end{equation}

{}From Eqs. (A1) and (A2) (with $N_{\alpha}(q')$ given
by the ground-state distribution $(24)$) we find that
the first derivatives $d R_{\alpha}^{(0)}(q)\over
{dq}$ can be expressed in terms of the functions
$R_{\alpha}^{(0)}(q)$ as follows

\begin{equation}
{d K^{(0)}(q)\over {dq}} =
{1\over {1 + {1\over {\pi}}{\cos K^{(0)}(q)\over {u}}
\int_{q_{Fs}^{(-)}}^{q_{Fs}^{(+)}}dq'
{1 \over {1 + [S^{(0)}(q') - {\sin K^{(0)}(q)\over {u}}]^2}}}}
\, ,
\end{equation}
and

\begin{equation}
{d S^{(0)}(q)\over {dq}} =
{1\over {{1\over {\pi}}\int_{q_{Fc}^{(-)}}^{q_{Fc}^{(+)}}dq'
{1 \over {1 + [S^{(0)}(q) - {\sin K^{(0)}(q')\over {u}}]^2}}
- {1\over {2\pi}}\int_{q_{Fs}^{(-)}}^{q_{Fs}^{(+)}}dq'
{1 \over {1 + [{1\over 2}(S^{(0)}(q) - S^{(0)}(q'))]^2}}}}
\, .
\end{equation}

{}From the same equations we find that the
second derivatives $d^2 R_{\alpha}^{(0)}(q)\over {dq^2}$
can be expressed in terms of the functions $R_{\alpha}^{(0)}(q)$
and its first derivatives (A10) and (A11) and read

\begin{eqnarray}
{d^2 K^{(0)}(q)\over {dq^2}} & = & [{d K^{(0)}(q)\over {dq}}]^3\{
{1\over {\pi}}{\sin K^{(0)}(q)\over {u}}
\int_{q_{Fs}^{(-)}}^{q_{Fs}^{(+)}}dq'
{1 \over {1 + [S^{(0)}(q') -
{\sin K^{(0)}(q)\over {u}}]^2}}\nonumber\\
& - & {2\over {\pi}}[{\cos K^{(0)}(q)\over {u}}]^2
\int_{q_{Fs}^{(-)}}^{q_{Fs}^{(+)}}dq'
{[S^{(0)}(q') - {\sin K^{(0)}(q)]\over {u}} \over
{\{1 + [S^{(0)}(q') - {\sin K^{(0)}(q)\over {u}}]^2\}^2}}\}
\, ,
\end{eqnarray}
and

\begin{eqnarray}
{d^2 S^{(0)}(q)\over {dq^2}} & = & [{d S^{(0)}(q)\over {dq}}]^3
\{{2\over {\pi}}\int_{q_{Fc}^{(-)}}^{q_{Fc}^{(+)}}dq'
{[S^{(0)}(q) - {\sin K^{(0)}(q')\over {u}}]\over
{\{1 + [S^{(0)}(q) - {\sin K^{(0)}(q')\over {u}}]^2\}^2}}
\nonumber\\
& - & {1\over {2\pi}}\int_{q_{Fs}^{(-)}}^{q_{Fs}^{(+)}}dq'
{[S^{(0)}(q) - S^{(0)}(q')]\over {\{1 + [{1\over 2}(S^{(0)}(q)
- S^{(0)}(q'))]^2\}^2}}\}
\, .
\end{eqnarray}

Let us introduce the functions $\widetilde{Q}_c^{(1)}(k)$
and $\widetilde{Q}_s^{(1)}(v)$ such that

\begin{equation}
Q_c^{(1)}(q) = \widetilde{Q}_c^{(1)}(K^{(0)}(q)) \, ,
\hspace{1cm}
Q_s^{(1)}(q) = \widetilde{Q}_s^{(1)}(S^{(0)}(q)) \, .
\end{equation}
Since we will often consider $k$ and $v$ integrations
instead of $q$ integrations, it is useful to
define the following limiting values

\begin{equation}
Q^{(\pm)} = K^{(0)}(q_{Fc}^{(\pm)}) \, ,
\hspace{1cm}
B^{(\pm)} = S^{(0)}(q_{Fs}^{(\pm)}) \, ,
\end{equation}
which refer to the pseudo-Fermi points $(12)-(14)$.
However, in order to be consistent with the order of
the expressions below, these values are to be replaced
by the corresponding leading-order terms $\pm Q$
and $\pm B$, respectively, which read

\begin{equation}
\pm Q = K^{(0)}(\pm q_{Fc}) \, ,
\hspace{1cm}
\pm B = S^{(0)}(\pm q_{Fs}) \, ,
\end{equation}
where the pseudo-Fermi points $\pm
q_{F\alpha}$ are defined in Eq. $(15)$. Introducing both
the distributions

\begin{equation}
N_{\alpha}(q) = N_{\alpha}^0(q) + \delta N_{\alpha}(q) \, ,
\end{equation}
and the first-order functions (A6) into Eqs. (A1) and (A2),
we find after expanding to first order that the functions
$Q_c^{(1)}(q)$ and $Q_s^{(1)}(q)$ are given by Eq. (A14) with
$\widetilde{Q}_c^{(1)}(k)$ and $\widetilde{Q}_s^{(1)}(v)$
defined by the following system of coupled integral
equations

\begin{equation}
\widetilde{Q}_c^{(1)}(k) = {1\over {\pi}}
\int_{q_s^{(-)}}^{q_s^{(+)}}dq' \delta N_s(q')
\tan^{-1}(S^{(0)}(q') - {\sin k\over {u}})
+ {1\over {\pi}}\int_{-B}^{B}
dv'{\widetilde{Q}_s^{(1)}(v')\over {1 +
[v'-{\sin k\over {u}}]^2}} \, ,
\end{equation}
and

\begin{eqnarray}
\widetilde{Q}_s^{(1)}(v) & = & - {1\over {\pi}}
\int_{q_c^{(-)}}^{q_c^{(+)}}dq' \delta N_c(q')
\tan^{-1}(v - {\sin K^{(0)}(q')\over {u}})\nonumber\\
& + & {1\over {\pi}}\int_{q_s^{(-)}}^{q_s^{(+)}}dq' \delta N_s(q')
\tan^{-1}({1\over 2}[v - S_0(q')])
+ {1\over {\pi}}\int_{-Q}^{Q}
dk'{\cos k'\over u}{\widetilde{Q}_c^{(1)}(k')\over {1 +
[v -{\sin k'\over {u}}]^2}}\nonumber\\
& - & {1\over {2\pi}}\int_{-B}^{B}
dv'{\widetilde{Q}_s^{(1)}(v')\over {1 +
[{1\over 2}(v - v')]^2}}
\, .
\end{eqnarray}
The use of Eq. (A18) in Eq. (A19) allows the introduction of
a single integral equation for $\widetilde{Q}_s^{(1)}(v)$.
Combining the obtained equations with Eqs. (A14) and
the phase-shift equations $(23)-(38)$ of Ref. \cite{Carmelo92b}
leads to expressions $(38)$.

In order to evaluate the second-order functions
$Q_c^{(2)}(q)$ and $Q_s^{(2)}(q)$ of the RHS of
Eq. (A4) we introduce the functions (A6) and
(A7) and distributions (A17) in Eqs. (A1) and (A2).
Expanding to second order we find after some algebra

\begin{equation}
Q_{\alpha}^{(2)}(q) = Q_{\alpha}^{(2,*)}(q) +
{1\over 2} {d\over {dq}}[[Q_{\alpha}^{(1)}(q)]^2] \, ,
\end{equation}
where

\begin{equation}
Q_{\alpha}^{(2,*)}(q) =
\widetilde{Q}_{\alpha}^{(2,*)}(R_{\alpha}^{(0)}(q)) \, ,
\end{equation}
and the functions $\widetilde{Q}_c^{(2,*)}(k)$ and
$\widetilde{Q}_s^{(2,*)}(v)$ are defined by the
following system of coupled integral equations

\begin{eqnarray}
\widetilde{Q}_c^{(2,*)}(k) & = & {1\over {\pi}}
\int_{q_s^{(-)}}^{q_s^{(+)}}dq' \delta N_s(q')
{d S^{(0)}(q')\over {dq'}}{Q_s^{(1)}(q')
\over {1 + [S^{(0)}(q') - {\sin k\over {u}}]^2}}\nonumber\\
& + & {1\over {2\pi}}[{d S^{(0)}(q)\over {dq}}]|_{q=q_{Fs}}
\sum_{j=\pm 1}{[Q_s^{(1)}(jq_{Fs})]^2
\over {1 + [jB - {\sin k\over {u}}]^2}}\nonumber\\
& + & {1\over {\pi}}\int_{-B}^{B}
dv'{\widetilde{Q}_s^{(2,*)}(v')\over {1 +
[v'-{\sin k\over {u}}]^2}} \, ,
\end{eqnarray}
and

\begin{eqnarray}
\widetilde{Q}_s^{(2,*)}(v) & = & {1\over {\pi}}
\int_{q_c^{(-)}}^{q_c^{(+)}}dq' \delta N_c(q')
{d K^{(0)}(q')\over {dq'}}{\cos K^{(0)}(q')\over u}
{Q_c^{(1)}(q')\over{1 + [v - {\sin K^{(0)}(q')\over
{u}}]^2}}\nonumber\\
& - & {1\over {2\pi}}\int_{q_s^{(-)}}^{q_s^{(+)}}dq'
\delta N_s(q'){d S^{(0)}(q')\over {dq'}}{Q_s^{(1)}(q')\over
{1 + [{1\over 2}(v - S_0(q'))]^2}}\nonumber\\
& + & {1\over {2\pi}}[{d K^{(0)}(q)\over
{dq}}]|_{q=q_{Fc}}{\cos Q\over u}
\sum_{j=\pm 1}{[Q_c^{(1)}(jq_{Fc})]^2\over
{1 + [v - {j\sin Q\over {u}}]^2}}\nonumber\\
& - & {1\over {4\pi}}[{d S^{(0)}(q)\over {dq}}]|_{q=q_{Fs}}
\sum_{j=\pm 1}{[Q_s^{(1)}(jq_{Fs})]^2\over
{1 + [{1\over 2}(v - jB)]^2}}\nonumber\\
& + & {1\over {\pi}}\int_{-Q}^{Q}
dk'{\cos k'\over u}{\widetilde{Q}_c^{(2,*)}(k')\over {1 +
[v -{\sin k'\over {u}}]^2}}\nonumber\\
& - & {1\over {2\pi}}\int_{-B}^{B}
dv'{\widetilde{Q}_s^{(2,*)}(v')\over {1 +
[{1\over 2}(v - v')]^2}} \, ,
\end{eqnarray}
respectively. Note that the free terms involve the
first-order functions.

In order to derive the first-order and second-order Hamiltonian
terms of Eqs. $(40)$ and $(41)$, we again consider eigenvalues
and deviations. We introduce in the Hamiltonian expression $(30)$
Eqs. (A6), (A7), and (A17) and expand the obtained expression.
Besides simpler terms, the first-order expression includes the
term $\int_{-Q}^{Q}dk[2t \sin k]\widetilde{Q}^{(1)}(k)$.
After some algebra we find the first-order term of Eq.
$(40)$ with the pseudoparticle bands given by expressions
$(19)-(21)$ of Ref. \cite{Carmelo91b}. Expanding $(30)$
to second-order leads to the following energy

\begin{eqnarray}
E_2 & = & {N_a\over {2\pi}}\{
\int_{q_c^{(-)}}^{q_c^{(+)}} dq' \delta N_c(q')
{d K^{(0)}(q')\over {dq'}}[2t\sin K^{(0)}(q')]Q_c^{(1)}(q')
\nonumber\\
& + & {1\over {2\pi}}[{d K^{(0)}(q)\over {dq}}]|_{q=q_{Fc}}
[t\sin Q]\sum_{j=\pm 1}[Q_c^{(1)}(jq_{Fc})]^2
+ \int_{-Q}^{Q}dk[2t \sin k]\widetilde{Q}_c^{(2,*)}(k)\}
\, ,
\end{eqnarray}
such that $\hat{H}^{(2)}|\eta_z,S_z\rangle =
E_2|\eta_z,S_z\rangle$. Inserting the suitable functions in
the RHS of Eq. (A24), performing some integrations by using
symmetry properties of the integral equations (A22) and
(A23), and replacing deviations by pseudomomentum normal-ordered
operators $(25)$ we find after some algebra expression $(41)$.
This can be rewritten in terms of the $f$ functions $(42)$
as given in the RHS of Eq. $(40)$.


\bigskip
$*$ Permanent address,
Department of Physics, University of \'Evora, Apartado 94, P-7001
\'Evora codex, Portugal.


\begin{references}
\bibitem[1]{Bethe}
        This ansatz was introduced for the case of the
        single-component ($\nu=1$) isotropic Heisenberg
        chain by H. A. Bethe, Z. Phys. {\bf 71}, 205 (1931).
\bibitem[2]{Yang}
        For one of the first generalizations of the Bethe
        ansatz to multicomponent $(\nu=2)$ systems see
        C. N. Yang, Phys. Rev. Lett. {\bf 19}, 1312
        (1967).
\bibitem[3]{Thacker}
        For a classical review of the Bethe ansatz and integrability
        in field theory and statistical mechanics see H. B. Thacker,
        Rev. Mod. Phys. {\bf 53}, 253 (1981).
\bibitem[4]{Lieb}
        Elliott H. Lieb and F. Y. Wu, Phys. Rev. Lett. {\bf 20},
        1445 (1968).
\bibitem[5]{Korepinrev}
        For a modern and comprehensive discussion of these issues,
        see V. E. Korepin, N. M. Bogoliubov, and A. G. Izergin,
        {\it Quantum Inverse Scattering Method and Correlation Functions}
        (Cambridge University Press, 1993).
\bibitem[6]{Ander}
        Philip W. Anderson, Phys. Rev. Lett. {\bf 64},
        1839 (1990); {\bf 65} 2306 (1990);
        P. W. Anderson and Y. Ren, in {\it High Temperature
        Superconductivity}, edited by K. S. Bedell,
        D. E. Meltzer, D. Pines, and J. R. Schrieffer
        (Addison-Wesley, Reading, MA, 1990).
\bibitem[7]{Haldane}
        F. D. M. Haldane, J. Phys. C {\bf 14}, 2585 (1981).
\bibitem[8]{Izergin}
        A.G. Izergin, V.E. Korepin, and N. Yu Reshetikhin,
        J. Phys. A: Math. Gen. {\bf 22}, 2615 (1989).
\bibitem[9]{Woy}
        F. Woynarovich, J. Phys. A {\bf 22}, 4243 (1989).
\bibitem[10]{Parola}
        Alberto Parola and Sandro Sorella,
        Phys. Rev. Lett. {\bf 64}, 1831 (1990).
\bibitem[11]{Schulz}
        H. J. Schulz, Phys. Rev. Lett. {\bf 64}, 2831
        (1990).
\bibitem[12]{Ren}
        Yong Ren and P. W. Anderson, Carg\`ese Lectures,
        1990 (unpublished).
\bibitem[13]{Frahm}
        Holger Frahm and V. E. Korepin, Phys. Rev. B {\bf 42},
        10553 (1990); {\bf 43}, 5653 (1991).
\bibitem[14]{Carmelo92}
        J. M. P. Carmelo and P. Horsch,
        Phys. Rev. Lett. {\bf 68}, 871 (1992).
\bibitem[15]{Carmelo92c}
        J. M. P. Carmelo, P. Horsch, and A. A. Ovchinnikov,
        Phys. Rev. B {\bf 46}, 14 728 (1992).
\bibitem[16]{Yang89}
        C. N. Yang, Phys. Rev. Lett. {\bf 63}, 2144
        (1989).
\bibitem[17]{Korepin}
        Fabian H. L. Essler, Vladimir E. Korepin, and
        Kareljan Schoutens, Phys. Rev. Lett. {\bf 67},
        3848 (1991); Nucl. Phys. B {\bf 372}, 559 (1992).
\bibitem[18]{Neto}
        J. M. P. Carmelo and A. H. Castro Neto,
        Phys. Rev. Lett. {\bf 70}, 1904 (1993).
\bibitem[19]{Carmelo91}
        J. Carmelo and A. A. Ovchinnikov, J. Phys.: Condens.
        Matter {\bf 3}, 757 (1991).
\bibitem[20]{Carmelo91b}
        J. Carmelo, P. Horsch, P.-A. Bares, and A. A. Ovchinnikov,
        Phys. Rev. B {\bf 44}, 9967 (1991).
\bibitem[21]{Carmelo92b}
        J. M. P. Carmelo, P. Horsch, and A. A. Ovchinnikov,
        Phys. Rev. B {\bf 45}, 7899 (1992).
\bibitem[22]{Campbell}
        J. M. P. Carmelo, P. Horsch, D. K. Campbell, and
        A. H. Castro Neto, Phys. Rev. B {\bf 48}, 4200 (1993).
\bibitem[23]{Neto93}
        J. M. P. Carmelo, A. H. Castro Neto, and D. K. Campbell
        preprint (1993).
\bibitem[24]{Heilmann}
        O. J. Heilmann and E. H. Lieb, Ann. N. Y.
        Acad. Sci. {\bf 172}, 583 (1971).
\bibitem[25]{Lieb89}
        E. H. Lieb, Phys. Rev. Lett. {\bf 62},
        1201 (1989).
\bibitem[26]{Zhang}
        C. N. Yang and S. C. Zhang,
        Mod. Phys. Lett. B {\bf 4}, 759 (1990).
\bibitem[27]{Ostlund}
        Stellan \"Ostlund, Phys. Rev. Lett. {\bf 69},
        1695 (1992).
\bibitem[28]{Essler}
        Fabian H. L. Essler and Vladimir E. Korepin,
        preprint ITP-SB-93-40 (1993) (to be published);
        preprint ITP-SB-93-45 and BONN-HE-93-39 (1993).
\bibitem[29]{Two}
        J. M. P. Carmelo, D. K. Campbell, A. H. Castro Neto,
        and N. M. R. Peres, preprint (1994).
\bibitem[30]{Nuno}
        J. M. P. Carmelo and N. M. R. Peres,
        preprint (1994).
\bibitem[31]{Castro}
        Walter Metzner and Carlo Di Castro,
        Phys. Rev. B {\bf 47}, 16107 (1993).
\bibitem[32]{Nuno94}
        J. M. P. Carmelo and N. M. R. Peres,
        unpublished (1994).
\bibitem[33]{Horsch94}
        J. M. P. Carmelo, P. Horsch, and N. M. R. Peres,
        unpublished (1994).
\bibitem[34]{Carmelo94}
        J. M. P. Carmelo, A. H. Castro Neto,
        D. K. Campbell, and N. M. R. Peres, preprint (1994).
\bibitem[35]{Yang69}
        C. N. Yang and C. P. Yang,
        J. Math. Phys. {\bf 10}, 1115 (1969).
\bibitem[36]{Pines}
        D. Pines and P. Nozi\`eres, in {\em The Theory of
        Quantum Liquids},
        (Addison-Wesley, Redwood City, 1966 and 1989), Vol. I.
\bibitem[37]{Baym}
        Gordon Baym and Christopher J. Pethick, in
        {\em Landau Fermi-Liquid Theory Concepts and Applications},
        (John Wiley \& Sons, New York, 1991).
\bibitem[38]{Korepin79}
        V. E. Korepin, Teor. Mat. Fiz. {\bf 41}, 169
        (1979) [Theor. Math. Phys. {\bf 41}, 953 (1980)].
\bibitem[39]{Zamo}
        Alexander B. Zamolodchikov and Alexey B. Zamolodchikov,
        Nucl. Phys. B {\bf 133}, 525 (1978).
\bibitem[40]{Andrei}
        N. Andrei and J. H. Lowenstein, Phys. Lett.
        {\bf 91B}, 401 (1980).
\bibitem[41]{Varma}
        C. M. Varma, P. B. Littlewood, S. Schmitt-Rink,
        E. Abrahams, and A. E. Ruckenstein,
        Phys. Rev. Lett. {\bf 63}, 1996 (1989).
\bibitem[42]{Guinea}
        F. Guinea, E. Louis, and J. A. Verg\'es,
        Phys. Rev. B {\bf 45}, 4752 (1992).
\bibitem[43]{Essler92}
        Fabian H. L. Essler, Vladimir E. Korepin, and
        Kareljan Schoutens, Phys. Rev. Lett.
        {\bf 68}, 2960 (1992).
\bibitem[44]{Hua}
        Hua Chen and Daniel Mattis, preprint University of
        Utah (1992).
\bibitem[45]{Pouget}
        J. P. Pouget, S. K. Khanna, F. Denoyer, R. Com\`es, A. F.
        Garito, and A. J. Heeger, Phys. Rev. Lett. {\bf 37},
        437 (1976).
\bibitem[46]{Jacob}
        C. S. Jacobsen, Ib Johannsen, and K. Bechgaard,
        Phys. Rev. Lett. {\bf 53}, 194 (1984).
\end{references}
\end{document}